\documentclass{article}

\usepackage{arxiv}

\usepackage[utf8]{inputenc} 
\usepackage[T1]{fontenc}    
\usepackage{hyperref}       
\usepackage{url}            
\usepackage{booktabs}       
\usepackage{amsmath}
\usepackage{amsfonts}       
\usepackage{amssymb}
\usepackage{nicefrac}       
\usepackage{microtype}      
\usepackage{lipsum}		
\usepackage{graphicx}
\usepackage{natbib}
\usepackage{doi}

\usepackage{epsfig}
\usepackage{ulem}
\usepackage{color}
\usepackage{dcolumn}
\usepackage{bm}
\usepackage{xcolor}
\usepackage{braket}
\usepackage{xfrac}

\title{Influence of transition mutations and disorder \\ on charge localization and transfer \\ along B-DNA sequences}

\date{} 					

\author{ {P. Banev, A. Falliera, and C. Simserides *} \\
National and Kapodistrian University of Athens, Department of Physics, \\
Panepistimiopolis, GR-15784, Zografos, Athens, Greece \\
\texttt{* csimseri@phys.uoa.gr}}

\renewcommand{\shorttitle}{\textit{Influence of transition mutations and disorder...} }

\begin{document}
\maketitle

\begin{abstract}
We illuminate the influence of transition mutations and disorder on charge localization and transfer along B-DNA sequences. Homopolymers are the best for charge transfer (cf. Refs.~ \cite{LVBMS:2018, MLTS:2019}). Hence, we consider as flawless a homopolymer sequence and then disturb it, introducing transition mutations and disorder. 
We exclude the possibility of charge transfer via the backbone that will be addressed soon in another work. We employ the Tight Binding (TB) Wire model to study the influence of transition mutations and the TB Fishbone Wire model to evaluate the influence of  disorder emanating either from the $\pi$ path or from the backbone. For the TB Wire parameters, we employ the parametrization created in Ref.~\cite{MLS:2023}, where another TB at atomic level was used, considering all valence orbitals of all atoms. We calculate the HOMO and LUMO regime eigenenergies and eigenvectors, the participation ratio (a measure of the localization of each eigenstate), the time-dependent probability to find the carrier at each site, the mean over time probability at each site, and the mean transfer rate from site to site.
Transition mutations increase localization in terms of participation ratio and impede charge transfer in terms of mean probability and transfer rates, provided the TB parameters involving mutated sites are significantly modified relative to the original. Disorder leads to severe modifications of participation ratios, i.e., increase of localization. Relevant changes occur on eigenenergies, mean probabilities at each site, and transfer rates.
\end{abstract}

\keywords{DNA \and charge transfer in biological systems \and mutations \and disorder}

\section{\label{sec:Introduction} Introduction}
In DNA double helix, \textit{normally}, a purine on one strand, adenine (A) or guanine (G), is connected via two or three hydrogen bonds with a complementary pyrimidine on the other strand, thymine (T) or cytosine (C), respectively. Each of the DNA bases A, G, T, C, forms towards the outer double helix, with one deoxyribose and one phosphate, a nucleotide. Usually, it is considered that charge movement proceeds via the so-called $\pi$ path of  interacting base pairs at the centre of the double helix. One could give thousands of citations, but let's limit ourselves to only a few \cite{EleySpivey:1962, Endres:2004, Schuster:2004I, Schuster:2004II}. Charge movement via the backbone has also been proposed \cite{Zhuravel:2020}. We will bypass this hypothesis excluding the possibility of charge movement via the backbone, but we will address it soon.

The \textit{central dogma of molecular biology} asserts that genetic information flows only in one direction: DNA to RNA to protein or RNA directly to protein. The flow of information from DNA to RNA is called \textit{transcription} and from RNA to protein \textit{translation}. DNA nucleotides are transcribed to their complementary RNA nucleotides and then translated in groups of three (triplets, called \textit{codons}) into amino acids that construct proteins. If a mutation occurs in DNA, e.g., if one base in a DNA triplet changes from one base to another, this will lead to a different RNA codon and hence to a different amino acid that will affect the proper protein production or it will lead to a \textit{start} or \textit{stop} codon that will crucially affect protein production. Different types of mutations found in a cell include \textit{point mutations} (change of one base to another), \textit{frame shift mutations} (insertion of a base in the sequence), \textit{non-sense mutations} (one that leads to a stop codon, that will stop production of the rest amino acids), and \textit{missense mutations} (change of one codon to another, but not to stop or start codon, hence, only the production of one amino acid will change). Since many codons lead to the same amino acid, missense mutations can be divided into \textit{silent mutations} (the change leads luckily to the same amino acid), \textit{conservative mutations} (the change leads to a new amino acid of the same type as the original) and \textit{non-conservative mutations} (the change leads to a new amino acid of different type to the original). Point mutations are divided in \textit{transitions}, i.e., interchange of purines, A $\leftrightarrow$ G or of pyrimidines, C $\leftrightarrow$ T, and \textit{transversions}, i.e., interchange purine $\leftrightarrow$ pyrimidine,~\cite{YangYoder:1999,Keller:2007,StoltzfusNorris:2015,Lyons:2017}. Other mutations involve proliferation of triplets, e.g. CAG repeats like in Huntington’s disease, Kennedy’s disease, Spinocerebellar ataxia 6, Spinocerebellar ataxia 7 \cite{MLS:2023} or CGG repeats, the sequence responsible for Fragile X syndrome~\cite{Nobile:2021, Mila:2018, Protic:2022, Dahlhaus:2018}, where the first diagonally located Cs of the repeated triplets get methylated.

In Ref.~\cite{MLS:2023} charge transport along ideal and natural DNA sequences as mutation detectors was studied, focusing on A-C mismatches and CAG repeats. \textit{Transport} means that an external driving force (a voltage) was applied: the experimentally relevant quantities are the current - voltage ($I$-$V$) curves. Here, we will examine charge \textit{transfer}, i.e., movement without external driving force: the experimentally relevant quantities are the transfer rates ($k$). We study charge transfer after oxidation (creation of a hole) or reduction (creation of an extra electron) in B-DNA homopolymers, including \textit{transition mutations}, within the Tight Binding (TB) Wire model, where each site is a base pair. As a benchmark, we deal with sequences G... (we write the 5'-3' direction), where the sites are G-C base pairs except for the mutated sites, which are mismatched A-C base pairs. We employ the parametrization of Ref.~\cite{MLS:2023} for the on-site energies of base pairs and the interaction integrals between base pairs. In Ref.~\cite{MLS:2023} another TB at the atomic level was used, considering all valence orbitals of all atoms. 

Disorder in DNA emanates from various sources. Here, we use the TB Fishbone Wire model to study it: the base pairs are the sites on the wire, but we also include sites at each deoxyribose, forming a fishbone. The aim is to study the influence of disorder on charge transfer again in homopolymers. Therefore, we allow for disorder on the on-site energies of base pairs and deoxyriboses and on the interaction parameters between base pairs as well as between base pairs and deoxyriboses: disorder emanates either from the $\pi$ path or from the backbone.

Today, as far as we know, there is no published work that uses two TB variants to show how transition mutations and disorder affect the electronic properties of a flawless DNA sequence and how they increase localization and impede charge transfer. We try to fill this gap studying an initially flawless system and the effects of mutations and disorder on that. This way it is easier to realize the effect of mutations and disorder. However, studies of mutation effects on the electronic properties of different DNA sequences, have been published in the past~\cite{Shih:2008, Shih:2012}. Ref.~\cite{Shih:2008} concerns point-mutation effects on charge transport properties of the tumor suppressor Gene p53 and Ref.~\cite{Shih:2012} concerns disease-related genes.
In Ref.~\cite{Shih:2011} the subject is charge transport in cancer-related genes and early carcinogenesis.

One transition mutation changes the on-site energy of the mutated base pair and the interaction integrals with the neighboring base pairs. This is clearly, plainly illustrated by the simple TB wire model. However, when we want to study disorder, we have to consider its various sources: from base pairs (on-site energies of base pairs and interaction integrals between base pairs), from the backbone (on-site energies of deoxyriboses), and from the interaction integrals between base pairs and deoxyriboses. The possibility of charge movement between backbone sites is neglected in the present work as we focus on that in a parallel work. Of course, the results of the TB Fishbone model converge to the results of the TB wire model if we extinguish the interaction integrals between base pairs and deoxyriboses, cf. Fig.~\ref{fig:WireVSFishboneWire}.

The article is organized as follows: In Sec. \ref{sec:Theory} we give a brief summary of the theory used. In Sec.~\ref{sec:ResultsDiscussion}
we present and discuss our results. Subsec.~
{\ref{subsec:W} refers to transitions within the Wire TB model and Subsec.~\ref{subsec:FW} refers to disorder within the Fishbone Wire TB model.

\section{\label{sec:Theory} Tight Binding (TB)}
The TB variants that we use have been extensively analyzed elsewhere, e.g., in Refs.~\cite{Simserides:2014, LVBMS:2018, MLTS:2019} (Wire Model) and in Ref.~\cite{SOML:2023} (Fishbone Wire Model). Therefore, we will refer readers to those articles for details. 
In the TB Wire Model the only sites we consider are the base pairs, while, in the TB Fishbone Wire Model we add one backbone site at each side of each base pair.
These two models are depicted schematically in Fig.~\ref{fig:WireVSFishboneWire}.
\begin{figure}[h]
\centering 
\includegraphics[width=0.6\textwidth]{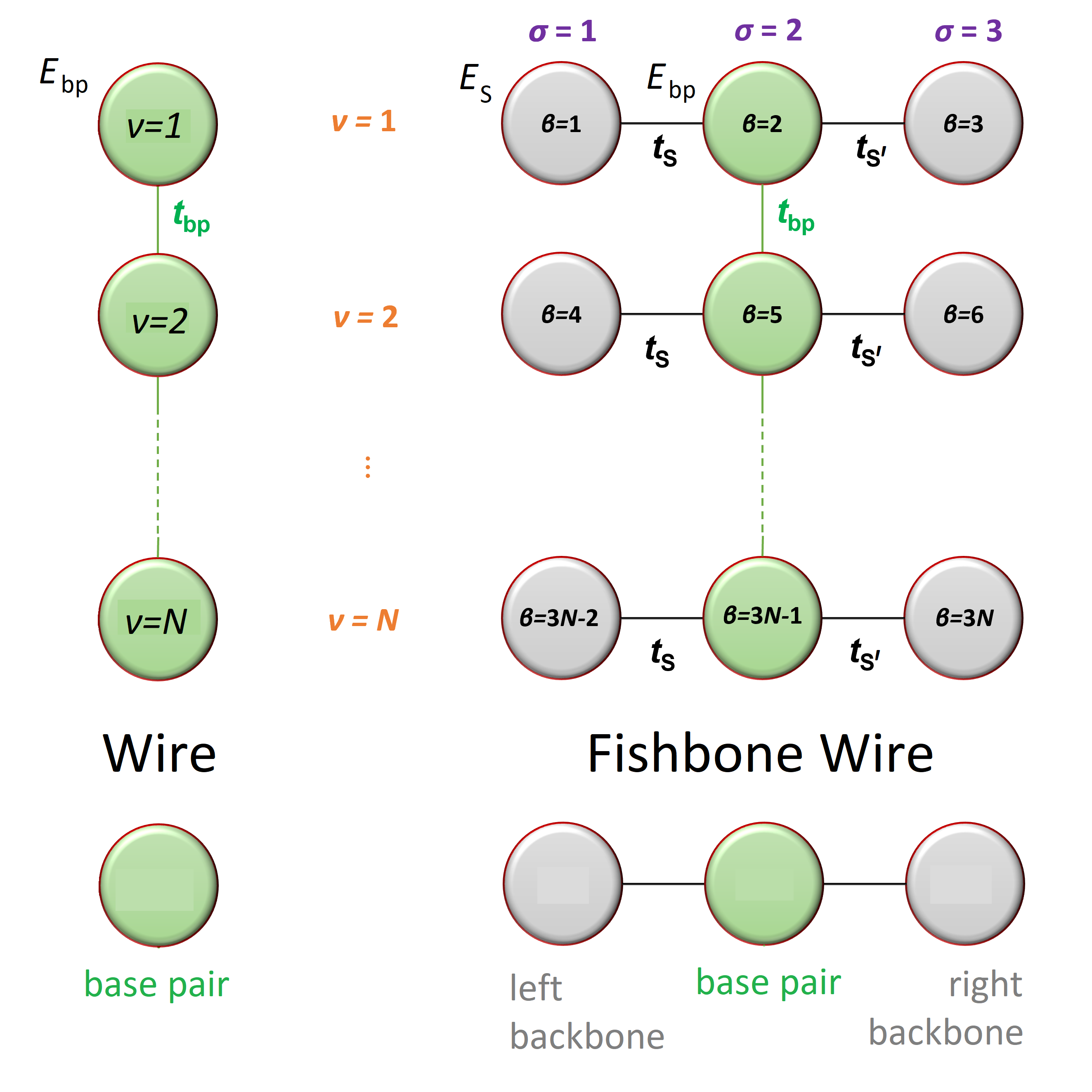}
\caption{Wire Model (left) vs. Fishbone Wire Model (right).
The results of the TB Fishbone model converge to the results of the TB wire model if we set $t_\mathrm{S} = 0 = t_\mathrm{S'}$.}
\label{fig:WireVSFishboneWire}
\end{figure}

In the Fishbone Wire Model, we name monomer a base pair together with its left and right backbone sites. In this work, we identify these backbone sites with deoxyriboses, since we know the deoxyribose on-site energy, while, that of the phosphate group is unknown to us, at least experimentally. Hence, here, each monomer has three sites: a base pair and two deoxyriboses connected to it. Between successive monomers, the only allowed interactions are those involving their respective base pairs: there are no interactions between backbone sites. We define three indices: $\sigma = 1, 2, 3$ is the strand index, $\nu=1,2, \dots N$ is the row or monomer index, and $\beta =1,2, \dots, 3N$ is the site index. The relation between these indices is  
\begin{equation}\label{eq:beta}
\beta= 3(\nu -1) + \sigma.
\end{equation}
Each base pair has its own on-site energy, $E_{\mathrm{bp}}$. The interaction integral between successive base pairs is denoted by $t_{\mathrm{bp}}$. We allow for different interaction integrals of the base pair with its left and right backbone site, $t_{\mathrm{S}}$ and $t_{\mathrm{S'}}$, respectively. The backbone (deoxyribose, sugar) on-site energy is denoted by $E_{\mathrm{S}}$.

The Hamiltonian describing the HOMO or the LUMO regime of a given DNA polymer within the Fishbone Wire Model is
\begin{align}\label{eq:HDNAfishbone}
\hat{H}  
&= \sum_{\nu=1}^{N}   E_{\nu,2} \ket{\nu,2} \bra{\nu,2}	 
+   \sum_{\nu=1}^{N-1} t_{\nu,2 \;,\; \nu\!+\!1,2} \ket{\nu,2} \bra{\nu\!+\!1,2} 
+   \sum_{\nu=2}^{N}   t_{\nu,2 \;,\; \nu\!-\!1,2} \ket{\nu,2} \bra{\nu\!-\!1,2} \nonumber \\
&+  \sum_{\nu=1}^{N} E_{\nu,1} \ket{\nu,1}\bra{\nu,1} 
+   \sum_{\nu=1}^{N} t_{\nu,2 \;,\; \nu,1} \ket{\nu,2} \bra{\nu,1} 
+   \sum_{\nu=1}^{N} t_{\nu,1 \;,\; \nu,2} \ket{\nu,1} \bra{\nu,2} \nonumber \\
&+  \sum_{\nu=1}^{N} E_{\nu,3} \ket{\nu,3} \bra{\nu,3} 
+   \sum_{\nu=1}^{N} t_{\nu,2 \;,\; \nu,3} \ket{\nu,2} \bra{\nu,3}
+   \sum_{\nu=1}^{N} t_{\nu,3 \;,\; \nu,2} \ket{\nu,3} \bra{\nu,2}.
\end{align}
$E_{\nu \, \sigma}$ and $t_{\nu \, \sigma \;,\; \nu^\prime \, \sigma^\prime}$ are the relevant on-site energies and interaction parameters, respectively. 
In the Wire Model, in the right hand side of Eq.~\ref{eq:HDNAfishbone}, we only keep the first three terms.

In brief, we assume that the time-independent (time-dependent) state of the sequence can be written as a linear combination of the HOMO or LUMO state of each site with time-independent (time-dependent) coefficients and solve the time-independent (time-dependent) Schr{\"o}dinger equation. Accordingly, we calculate quantities relevant to the electronic structure such as the HOMO and LUMO regime eigenenergies and eigenvectors and the participation ratio that is a measure of localization of each eigenstate \textit{as well as} quantities relevant to the temporal evolution after an initial oxidation (creation of a hole at a site's HOMO) or reduction (creation or an extra electron at a site's LUMO) such as the time-dependent probability to find the carrier at each site, mean over time probability at each site, and mean transfer rate from site to site. $N$ is the total number of base pairs and $M$ is the total number of sites. For the TB Wire Model, $M=N$; for the TB Fishbone Wire model, $M=3N$ because for each base pair we have one deoxyribose on its either side. The TB parameters that we use for the Wire Model are shown in Tables \ref{Table:OnsiteEnergies} and 
\ref{Table:InteractionIntegrals}. The TB interaction integrals between a base pair and its left and right deoxyribose are denoted by $t_\mathrm{S}$ and $t_\mathrm{S'}$, respectively. The on-site energy of  deoxyribose is denoted by $E_\mathrm{S}$.

\begin{table}[h]
\centering
\small
\caption{TB Wire Model. HOMO and LUMO on-site energies of base pairs. At G-C base pair, LUMO is of $\sigma^*$ character; we also give the (slightly higher)  $\pi^*$ character LUMO. We include the mismatched base pair created by the transition mutation G$\to$A, A-C (again LUMO is of $\sigma^*$ character; we also give the (slightly higher) $\pi^*$ character LUMO). This parametrization was produced by TB, using all atomic valence orbitals of all atoms \cite{MSF:2021,MLS:2023}. The mismatched base pair A-C is also called m.}
\label{Table:OnsiteEnergies}
\begin{tabular}{|c|c|c|} \hline
base pairs & $E^{\mathrm{bp}}_{\mathrm{H}}$ (eV) & $E^{\mathrm{bp}}_{\mathrm{L}}$ (eV) \\ \hline
G-C  & $-$8.30 ($\pi$) & $-$4.14 ($\pi^*$) \\ 
     &                 & $-$4.43 ($\sigma^*$) \\ \hline
A-T  & $-$8.49 ($\pi$) & $-$4.31 ($\pi^*$) \\ \hline
A-C  & $-$8.43 ($\pi$) & $-$4.23 ($\pi^*$) \\
or m &                 & $-$4.43 ($\sigma^*$) \\ \hline
\end{tabular}
\end{table}

\begin{table}[h]
	\centering
	\small
	\caption{TB Wire Model. The absolute values of the interaction parameters between consecutive base pairs, either between their HOMOs either between their LUMOs. Since at the G-C base pair, the LUMO is of type $\sigma^*$, when a G-C base pair is involved, we also give the parameters that correspond to the (slightly higher) $\pi^*$ type LUMO. This parametrization was produced by TB, using all atomic valence orbitals of all atoms \cite{MSF:2021,MLS:2023}. We include the interaction parameters, involving the base pair that results from transition mutation G$\to$A, i.e., A-C (for brevity we call it Am). Since at the base pair A-C, LUMO is of type $\sigma^*$, when an A-C base pair is involved, we also give the parameters referring to the (slightly higher) $\pi^*$ type LUMO. We give the absolute values of the interaction integrals since the spectrum of tridiagonal, irreducible, real, symmetric matrices (as all our matrices are, within the Wire Model) does not depend on the signs of their off-diagonal entries~\cite{MLTS:2019}. The mismatched A-C base pair is also called m.}
	\label{Table:InteractionIntegrals}
	\begin{tabular}{|c|c|c|} \hline
		base sequence  & $ t^{\mathrm{bp}}_\mathrm{H}$ (meV) ($\pi-\pi$) & $t^{\mathrm{bp}}_{\mathrm{L}}$ (meV) \\ \hline
		GG $\equiv$ CC & 116                      &  2 ($\pi^*-\pi^*$) \\ 
		&                          & 92 ($\sigma^*-\sigma^*$) \\ \hline
		AA $\equiv$ TT &  38                      & 22 ($\pi^*-\pi^*$)  \\ \hline
		GC             &  10                      & 19 ($\pi^*-\pi^*$) \\ 
		&                          &  2 ($\sigma^*-\sigma^*$) \\ \hline
		CG             &  75                      &  9 ($\pi^*-\pi^*$) \\ 
		&                          &  1 ($\sigma^*-\sigma^*$) \\ \hline  
		AT             &  50                      &  1 ($\pi^*-\pi^*$) \\ \hline
		TA             &  37                      &  2 ($\pi^*-\pi^*$) \\ \hline
		CT $\equiv$ AG &  37                      & 11 ($\pi^*-\pi^*$) \\ 
		&                          & 11 ($\sigma^*-\sigma^*$) \\ \hline
		TC $\equiv$ GA & 142                      &  6 ($\pi^*-\pi^*$) \\ 
		&                          &  3 ($\sigma^*-\sigma^*$) \\ \hline
		CA $\equiv$ TG &  28                      &  9 ($\pi^*-\pi^*$) \\ 
		&                          &  2 ($\sigma^*-\sigma^*$) \\ \hline
		AC $\equiv$ GT &  16                      &  1 ($\pi^*-\pi^*$) \\
		&                          &  1 ($\sigma^*-\sigma^*$) \\ \hline
		Gm            & 130                      &  8 ($\pi^*-\pi^*$) \\
		&                          & 89 ($\sigma^*$-$\sigma^*$) \\ \hline
		mG            &  31                      & 20 ($\pi^*-\pi^*$) \\   
		&                          & 90 ($\sigma^*$-$\sigma^*$) \\ \hline
		mm           &  36                      & 25 ($\pi^*-\pi^*$) \\
		&                          & 90 ($\sigma^*$-$\sigma^*$) \\ \hline
	\end{tabular} 
\end{table}

At this point, we would like to comment that DNA is susceptible to distortions of rotative and translative nature. Hence, variations of the TB parameters are expected in real situations. TB parameters have been obtained by different methods by different authors, cf. e.g., Ref.~\cite{Simserides:2014} and references therein. In Ref.~\cite{MMLSF:2021} Molecular Dynamics (MD) was used to estimate distortions; DFT and TB to evaluate their influence on the modification of on-site energies and interaction integrals. We comment on this issue below, analysing disorder. For a comparison of our TB methods with experiments, cf. e.g. Ref.~\cite{MLTS:2019}, where we compare with the experimentally deduced (by transient absorption spectroscopy) transfer rates of Ref.~\cite{Vura-Weis:2009,Conron:2010}.
As far as we know, there are no experiments on the influence of mutations and disorder on transfer rates, yet.

The mean transfer rate~\cite{Simserides:2014}, $k_{\alpha \beta}$, describes the rate of carrier transfer from site $\alpha$ to site $\beta$,  including both the amount of probability that is transferred and the time-scale of the phenomenon. It is a quantity defined for coherent charge transfer. For initial carrier placement at site $\alpha$ (i.e., site $\alpha$ is where initially the oxidation or reduction occurs), it is defined as 
\begin{equation}\label{eq:k}
k_{\alpha \beta} = \frac{\braket{|C_\beta(t)|^2}}{t_{\alpha\beta}},
\end{equation}
where $t_{\alpha \beta}$ is the time at which the time-dependent probability to find the carrier at site $\beta$, $|C_\beta(t)^2|$, becomes equal to its mean over time value, $\langle |C_\beta(t)^2|\rangle$, for the first time.

The participation ratio (PR)~\cite{Wegner:1980}, quantifies the localization of eigenstates, i.e., it shows at which extent an eigenstate is localized to one or more sites. If $M$ is the total number of sites, then the participation ratio of eigenstate $k$ is defined as
\begin{equation}
\mathrm{PR}_k = \frac{1}{{M \sum_{\alpha=1}^M {v_{\alpha k}}^4}},
\end{equation}
where $v_{\alpha k}$ is the site $\alpha$ component of eigenvector $k$, corresponding to the eigenenergy $E_k$. The eigenvectors are normalized, i.e.,
$\sum_{\alpha=1}^M{{v_{\alpha k}}^2}=1$, $\forall k$.

$\bullet$
\textit{For localization at one site} $\beta$, we have 
$v_{\beta k} = 1 \;  \& \;  v_{\alpha k} = 0,   \forall \alpha \neq \beta \; \Rightarrow \; \mathrm{PR}_k = \frac{1}{M}.$
Therefore, the asymptotic limit for an infinitely large system is zero, $\lim_{M \to \infty} \mathrm{PR}_k = 0.$

$\bullet$ \textit{At the other limit of equidistribution at all sites},
${v_{\alpha k}} = \frac{1}{\sqrt{M}}, \; \forall \alpha \; \Rightarrow \; \mathrm{PR}_k = 1.$

The participation ratio (PR)~\cite{Wegner:1980, Pruisken:1985, Carrillo-NunezSchulz:2008, Dey:2011} and similar quantities as the inverse participation ratio (IPR) have been used for the characterization  of localization of states in various areas of physics, materials science and related disciplines either in a quantum or in a classical context~\cite{Wegner:1980, Farkas:2001, Goltsev:2012, Pastor-SatorrasCastellano:2016, Safari:2017, Moretti:2019}.

\section{\label{sec:ResultsDiscussion} Results and Discussion}

\subsection{\label{subsec:W} Transitions, TB Wire model}
\begin{figure}[h]
\centering
\includegraphics[width=0.45\textwidth]{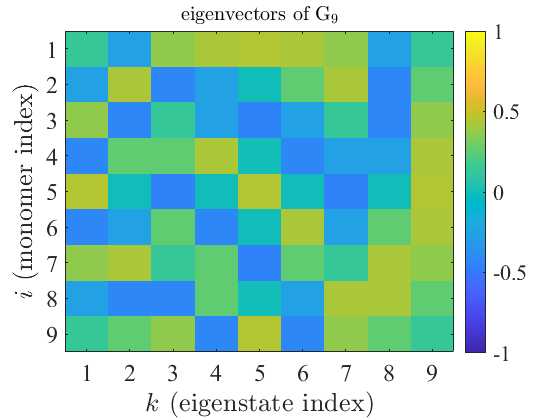}
\includegraphics[width=0.45\textwidth]{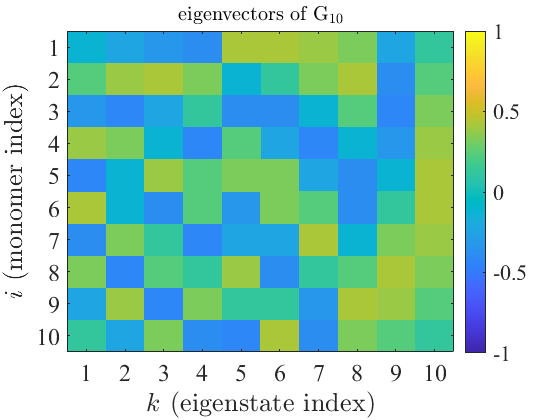}
\caption{TB Wire Model. The eigenvectors of G$_9$ and G$_{10}$ sequences. HOMO regime. Eigenstate index 1 (9 or 10) corresponds to the highest (lowest) eigenenergy.}
\label{fig:V_N=9and10_HOMO}
\end{figure}

The TB Wire Model parameters have been produced 
by another TB model which utilizes all valence orbitals of all atoms \cite{MSF:2021,MLS:2023}. They are given in Tables~\ref{Table:OnsiteEnergies} and
\ref{Table:InteractionIntegrals}. 

As a benchmark sequence we choose G$_9$ because it is purely symmetric in the sense that there exists a central monomer, the fifth. We also choose its even version, the G$_{10}$ sequence. Before including mutations, let us analyze the pure G$_9$ and G$_{10}$ sequences in terms of eigenvectors, $V_{ik}$, where $i$ is the monomer index and $k$ is the eigenstate index. These are depicted in Fig.~\ref{fig:V_N=9and10_HOMO}. Eigenstate index 1 (9 or 10) corresponds to the highest (lowest) eigenenergy. 

For comparison, we also depict the eigenvectors of the \textit{cyclic} versions of G$_9$ and G$_{10}$ sequences in Fig.~\ref{fig:V_N=9and10_HOMOcyclic}. \textit{Cyclic} means that we have connected the first and the last monomer letting them interact with the same interaction integral as between nearest neighbors. We observe that only the lowest eigenenergy of the cyclic case corresponds to equidistribution that will produce  PR = 1. 

\begin{figure}[h]
\centering
\includegraphics[width=0.45\textwidth]{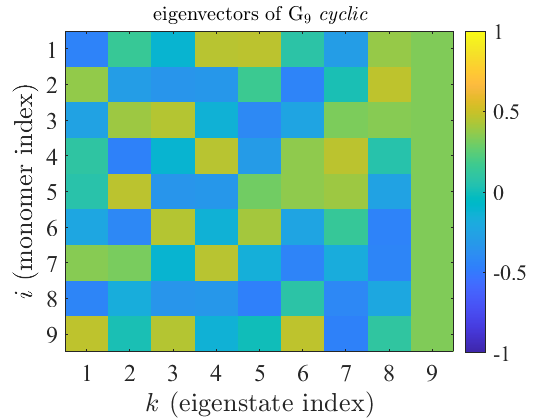}
\includegraphics[width=0.45\textwidth]{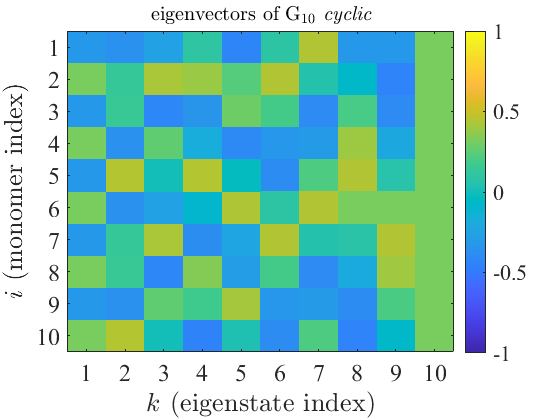}
\caption{TB Wire Model. The eigenvectors of G$_9$ and G$_{10}$ \textit{cyclic} sequences. 
HOMO regime. Eigenstate index 1 (9 or 10) corresponds to the highest (lowest) eigenenergy.}
\label{fig:V_N=9and10_HOMOcyclic}
\end{figure}

Insertion of one or two mutations in such sequences 
affects eigenvectors in the way shown in 
Fig.~\ref{fig:V_N=9and10_HOMO_1mutation} and 
Fig.~\ref{fig:V_N=9and10_HOMO_2mutations}, respectively.
We can now spot monomers with high absolute values of eigenvector components. Hence, localization is enhanced by including one or two mutations.

\begin{figure}[t]
\centering
\includegraphics[width=0.45\textwidth]{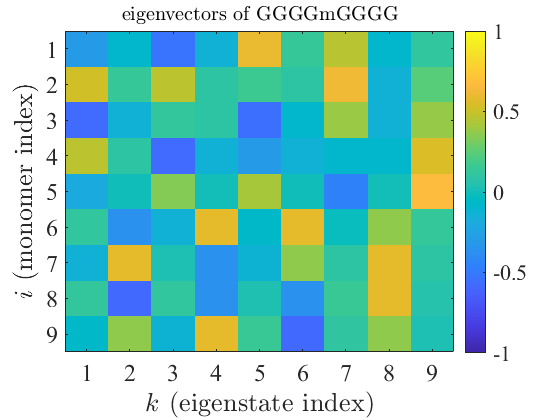}
\includegraphics[width=0.45\textwidth]{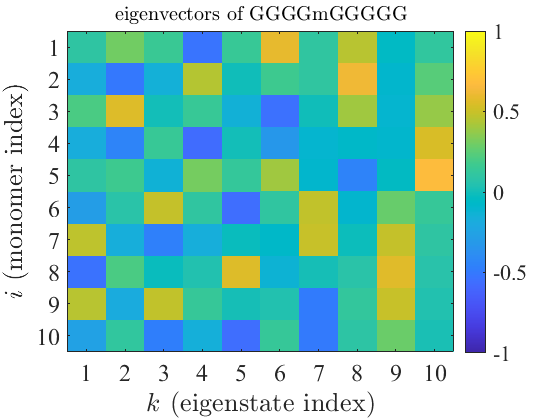}
\caption{TB Wire Model. The eigenvectors of the sequences GGGGmGGGG and GGGGmGGGGG. HOMO regime. Eigenstate index 1 (9 or 10) corresponds to the highest (lowest) eigenenergy.}
\label{fig:V_N=9and10_HOMO_1mutation}
\end{figure}

\begin{figure}[t]
\centering
\includegraphics[width=0.45\textwidth]{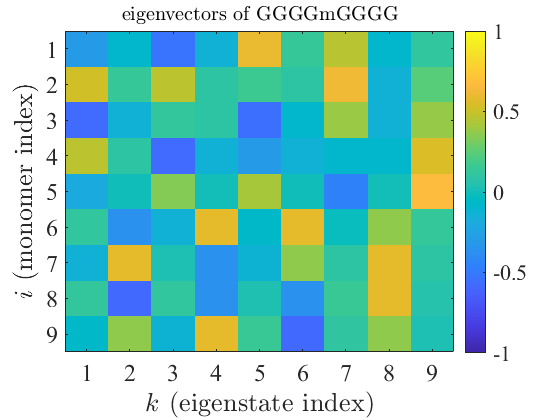}
\includegraphics[width=0.45\textwidth]{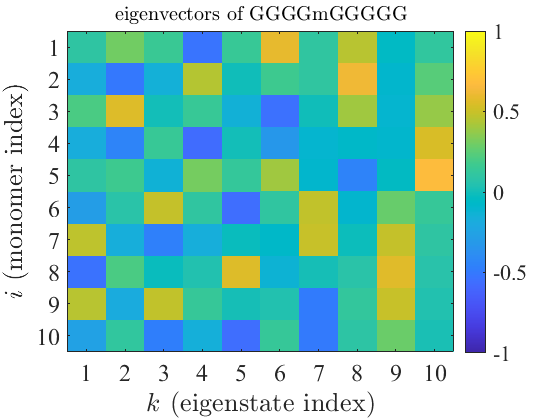}
\caption{TB Wire Model. The eigenvectors of sequences GmGGmGGGG and GmGGmGGGGG. HOMO regime. Eigenstate index 1 (9 or 10) corresponds to the highest (lowest) eigenenergy.}
\label{fig:V_N=9and10_HOMO_2mutations}
\end{figure}

The influence of one or two transitions G$\leftrightarrow$A on the PR which reflects changes in the eigenvectors (as an example for the HOMO regime) is shown in Fig.~\ref{fig:PRHOMO}. We observe that generally PR decreases as we introduce one and then two mutations, i.e., inserting A-C mismatches the system moves towards localization. 
In~Fig.~\ref{fig:PRHOMOofN}, we show the participation ratio, as we increase the length of the G... polymer: $\lim_{N \to \infty} \mathrm{PR}_{\textrm{G}...} = 2/3$. To be exact, this holds for even $N$. For odd $N$, there is one exception: the middle eigenenergy state (e.g., for $N=9$, the 5th) is more localized: it has non-negligible components only for odd monomers.

\begin{figure}[h]
\centering
\includegraphics[width=0.45\textwidth]{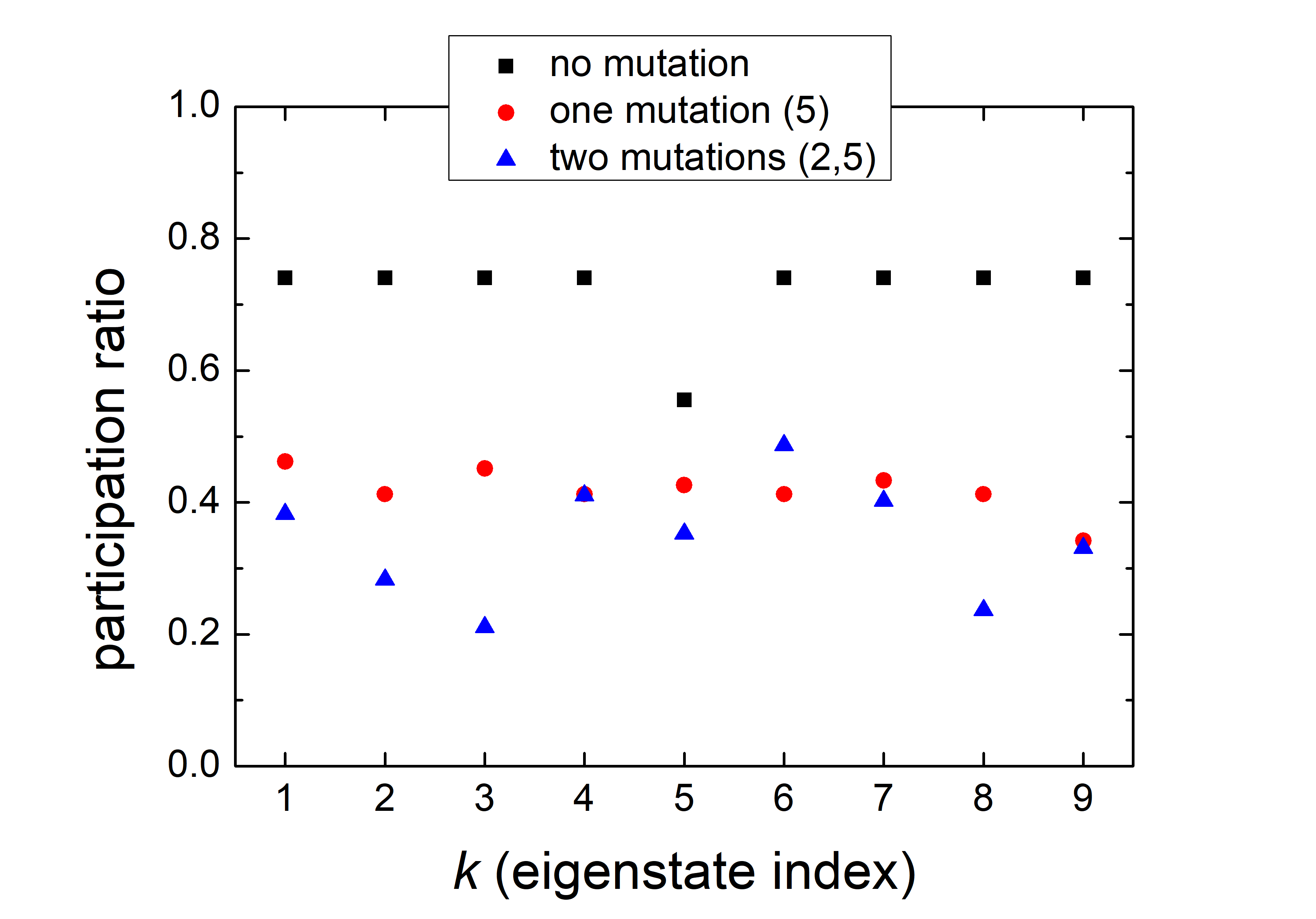}
\includegraphics[width=0.45\textwidth]{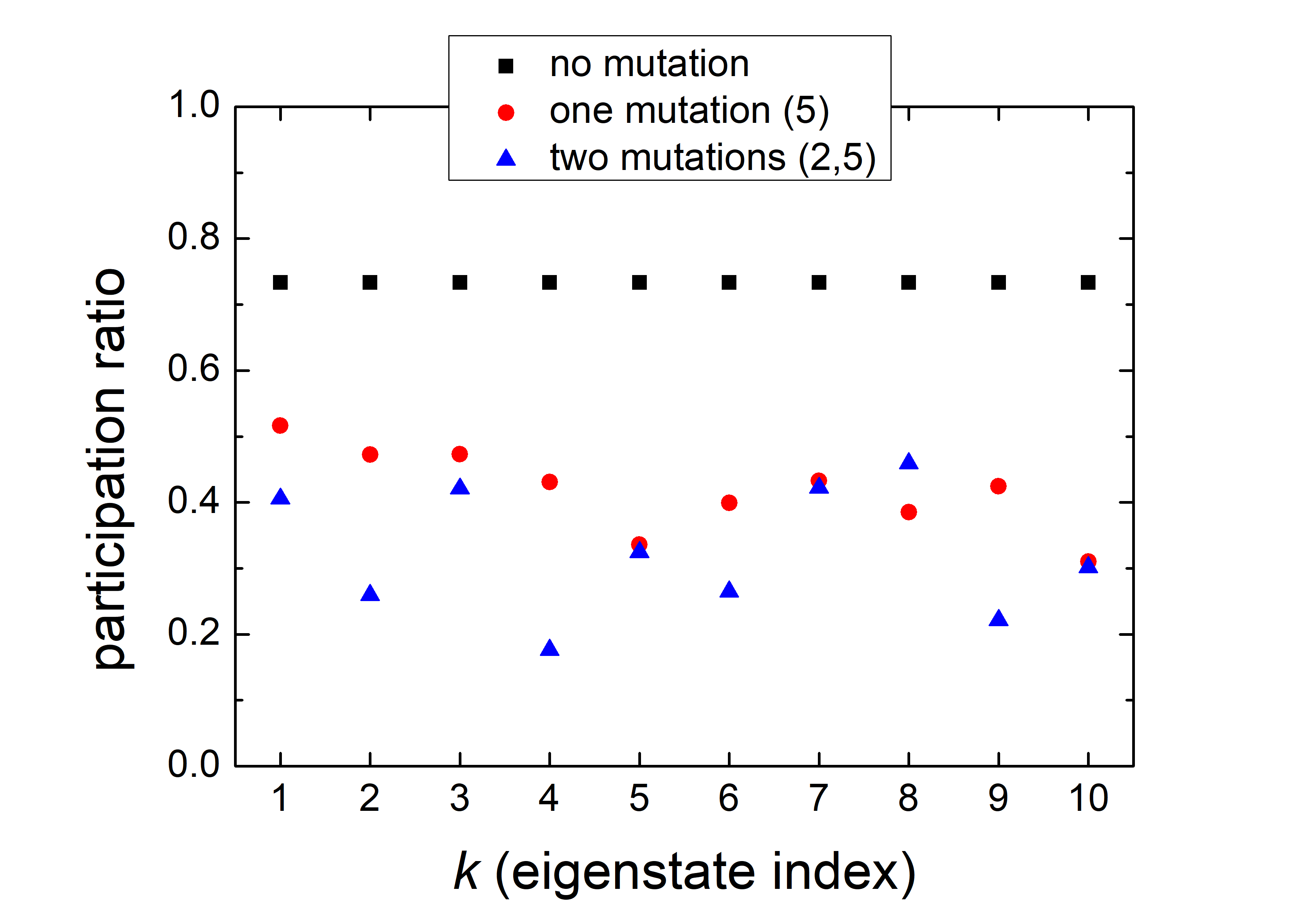}
\caption{TB Wire Model. Participation ratio (PR) for G... oligomers with $N=9$ and $N=10$, HOMO regime.}
\label{fig:PRHOMO}
\end{figure}

\begin{figure}[h]
\centering
\includegraphics[width=0.45\textwidth]{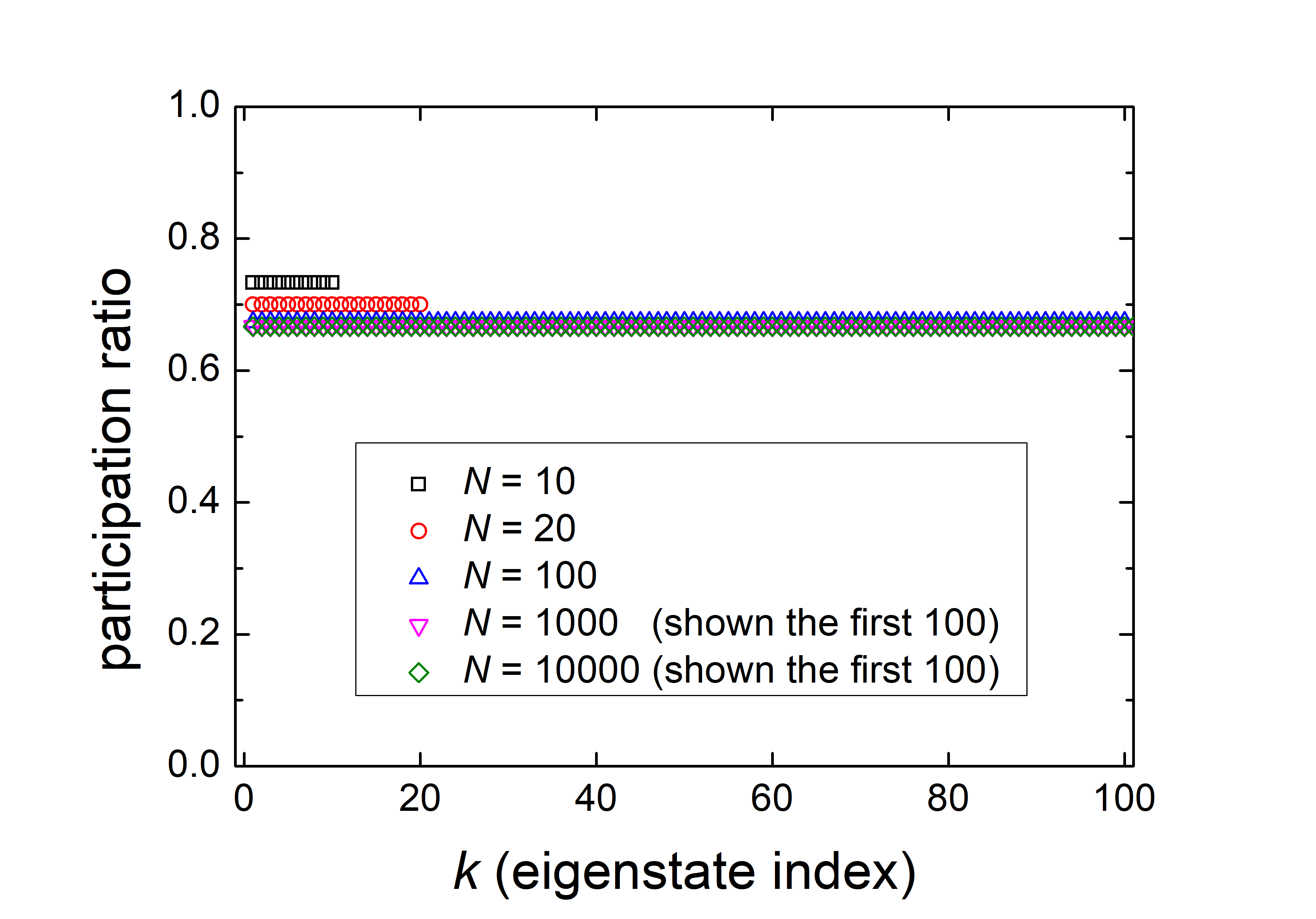}
\caption{TB Wire Model. Participation ratio, as we increase the length of the G... polymer, with $N$ even. $\lim_{N \to \infty} \mathrm{PR}_{\textrm{G}...} = 2/3$.}
\label{fig:PRHOMOofN}
\end{figure}

\begin{figure}
\centering
\includegraphics[width=0.45\textwidth]{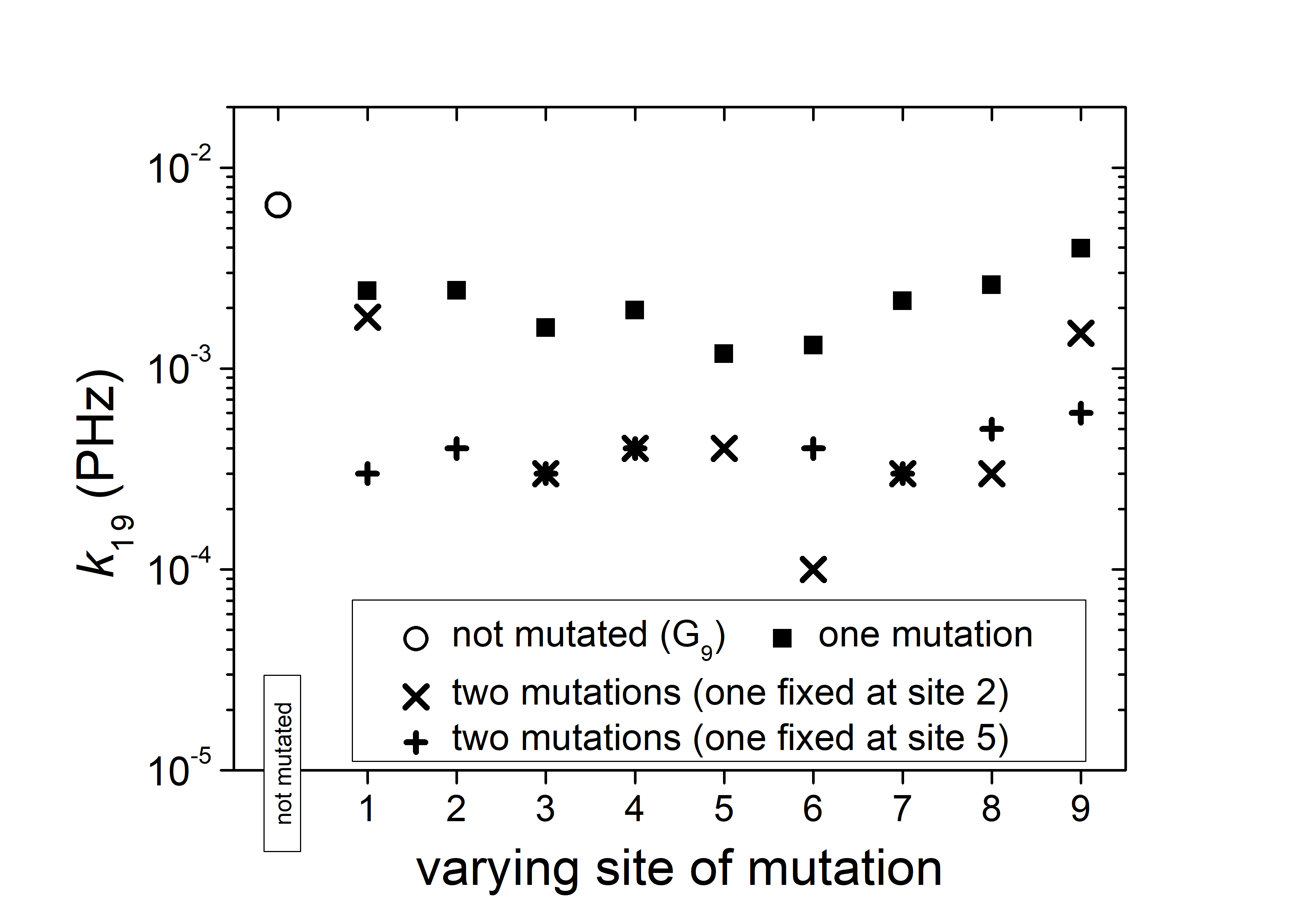} \\
\includegraphics[width=0.45\textwidth]{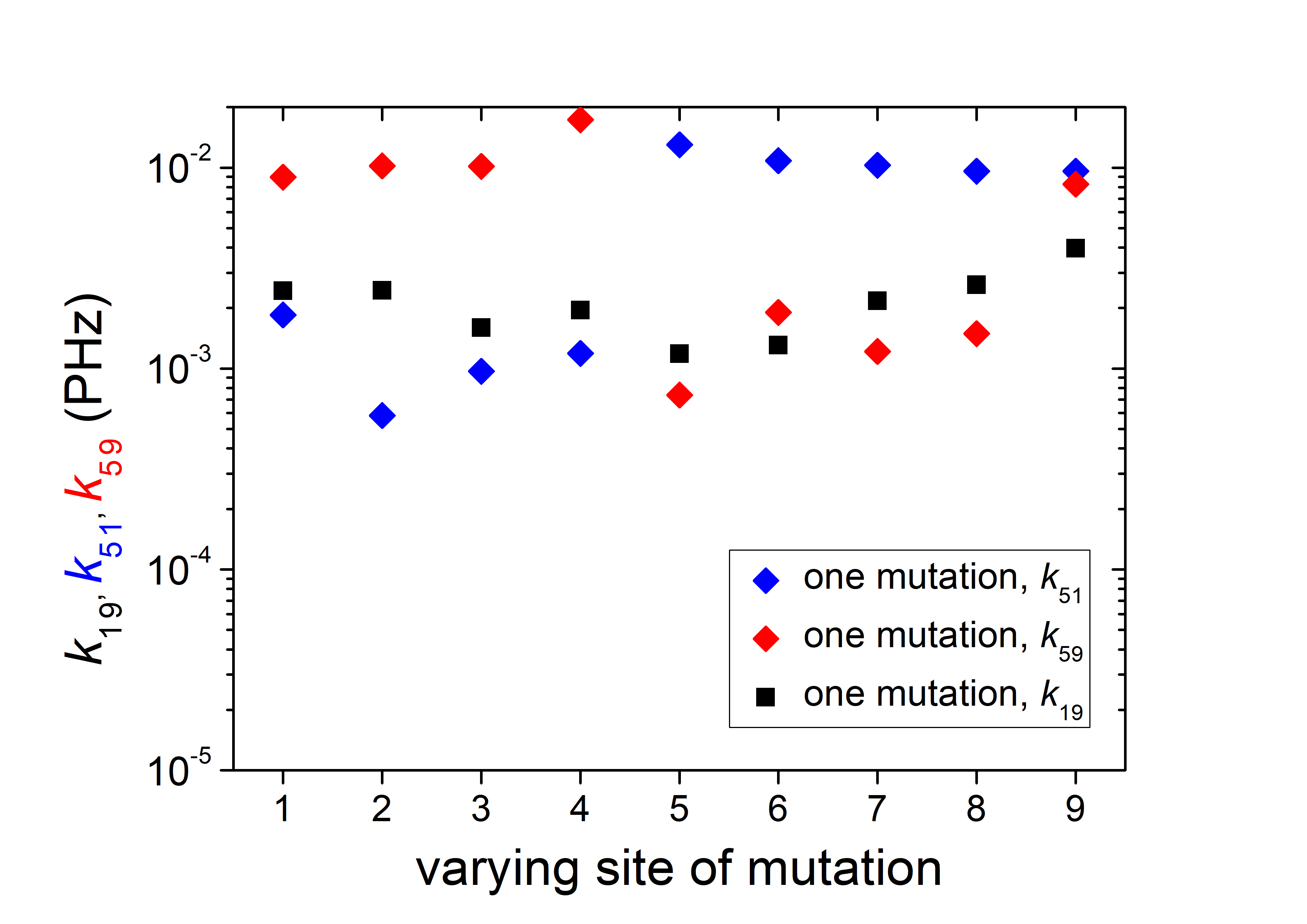} \\
\includegraphics[width=0.45\textwidth]{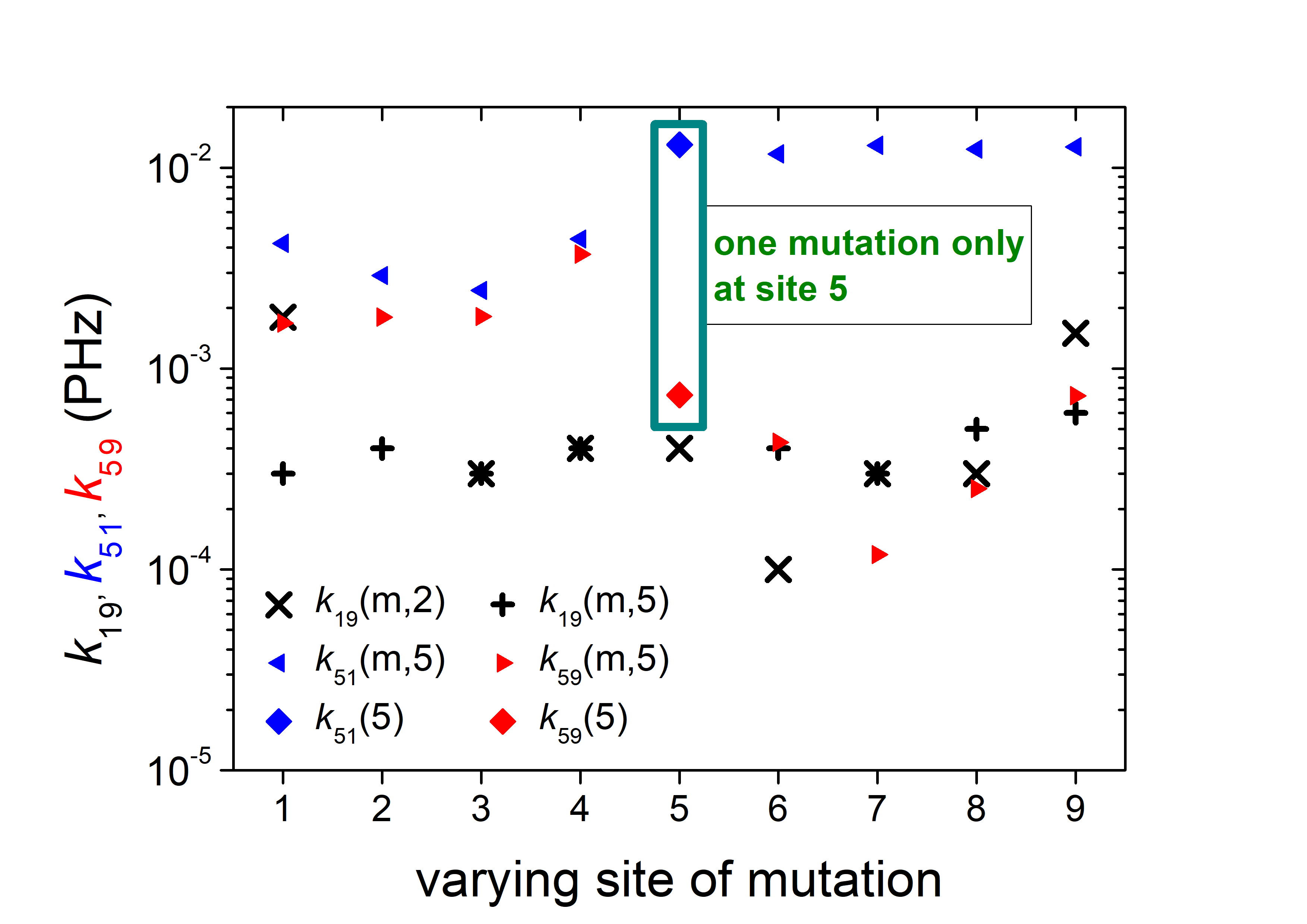}
\caption{TB Wire Model. \textit{Upper.} Transfer rates from site 1 (initial oxidation at one end) to site 9 (other end) for the not mutated G$_9$ sequence ($\circ$) as well as for one mutation of varying position ($\blacksquare$),
and two mutations, one fixed at site 2 and the other of varying position ($\times$) and one fixed at site 5 and the other of varying position ($+$). \textit{Middle.} A comparison of $k_{19}$ with the transfer rates from the middle monomer (5), where the initial oxidation now occurs, to the end monomers (1 and 9), $k_{51}$ and $k_{59}$, respectively, in a sequence with one mutation of varying position.
\textit{Lower.} 
$k_{19}$ with two mutations (one fixed at site 2) and
with two mutations (one fixed at site 5) vs.  
$k_{51}$ with two mutations (one fixed at site 5) and
$k_{59}$ with two mutations (one fixed at site 5).
$k_{51}$ with only one mutation fixed at site 5 and 
$k_{59}$ with only one mutation fixed at site 5 
are also shown, for comparison.}
\label{fig:kHOMO}
\end{figure}

The influence of one or two transitions G$\leftrightarrow$A on the eigenspectra of G$_9$ is shown in Fig.~\ref{fig:EigenSpectra}. Significant modifications occur only in the HOMO regime. This is because the relevant TB parameters remain almost identical in the LUMO regime (for this specific transition), as evident from  Tables~\ref{Table:OnsiteEnergies} and
\ref{Table:InteractionIntegrals}. Therefore, mutation(s) will have effect on charge transfer after oxidations but not after reductions. For this reason, below we will only show results for holes, i.e., what happens, after an initial oxidation, at the sites of sequence G$_9$, including one or two A-C mismatches. 

Figure~\ref{fig:MeanProbio1} (\ref{fig:MeanProbio5}) presents the probability to find the hole at each site, after an initial oxidation at site 1 (5), in the presence of one mutation of varying position in the sequence. 
We include the not mutated cases, for comparison, 
in Fig.~\ref{fig:MeanProbio1+io5}.
In Fig.~\ref{fig:MeanProbio1}, most of the probability remains before the mutated site, in all cases. Similarly, in Fig.~\ref{fig:MeanProbio5}, it is evident that the mutation hinders carrier transfer. 
The \textit{upper} panel of Fig.~\ref{fig:kHOMO} shows $k_{19}$, the transfer rate from site 1 (initial oxidation at one end of the sequence) to site 9 (the other end) for various cases. $k_{19}$ for the not mutated G$_9$ sequence is shown with $\circ$, for a sequence with one mutation of varying position with $\blacksquare$, for two mutations one fixed at site 2 and the other of varying position with $\times$ and for two mutations one fixed at site 5 and the other of varying position with $+$. We observe that even one mutation is enough to decrease the transfer rate significantly; two mutations reduce it $\approx$ by one more order of magnitude. The \textit{middle} panel of Fig.~\ref{fig:kHOMO} shows a comparison of $k_{19}$ with the transfer rates from the middle monomer (5), where the initial oxidation now occurs, to the end monomers (1 and 9), $k_{51}$ and $k_{59}$, respectively, in a sequence with one mutation of varying position. For $k_{51}$ and $k_{59}$ the carrier has to pass half of the oligomer, while for $k_{19}$ it has to pass the whole oligomer. Nevertheless, we observe that when there is a mutation in the path, the transfer rate decreases significantly. The \textit{lower} panel of Fig.~\ref{fig:kHOMO} shows
$k_{19}$ with two mutations (one fixed at site 2) and
with two mutations (one fixed at site 5) vs.  
$k_{51}$ with two mutations (one fixed at site 5) and
$k_{59}$ with two mutations (one fixed at site 5).
$k_{51}$ and $k_{59}$ with only one mutation fixed at site 5 are also shown, for comparison. Again, mutations obstruct charge transfer. 
Details on these calculations can be found in Ref.~\cite{Falliera:2024}.

\textit{Conclusion:} Mutations increase localization in terms of PR and impede charge transfer in terms of mean probability and transfer rates, provided that the TB parameters involving mutated sites are significantly modified relative to the original ones.

\subsection{\label{subsec:FW} Disorder, TB Fishbone Wire model}
We will evaluate the effect of disorder with the help of the TB Fishbone Wire Model~\cite{SOML:2023}. 
By $t_\mathrm{S}$ we denote the interaction integral between a G-C base pair and a deoxyribose (``sugar''), by $E_\mathrm{S}$ the deoxyribose on-site energy, which is around 9.0 eV, if taken equal to the absolute value of its ionization energy~\cite{Ghosh:2012}, $t_\mathrm{G}$ is a short for the interaction integral between successive G-C base pairs, and $E_\mathrm{G}$ is a short for the on-site energy of the G-C base pair. As a benchmark sequence we keep the pentamer G$_5$, because it is purely symmetric in the sense that there exists a central monomer, the third. The interaction parameters between deoxyriboses and base pairs can be calculated~\cite{Baumeier:2010}; the values we use are representative~\cite{Morphis:2023}; details will be published elsewhere.
For the Fishbone Wire Model we study a homopolymer of chain length 5, as an example, because it is easier to discuss results as it only has 15 sites. Similar results hold for longer chains, of course.

Before including disorder, let us analyze the pure G$_5$ sequence in terms of eigenvectors, $V_{ik}$, where $i$ is the site index and $k$ is the eigenstate index. Sites 2, 5, 8, 11, 14 are the G-C base pairs and the rest are deoxyriboses on either site of base pairs: sites 1, 4, 7, 10, 13 (3, 6, 9, 12, 15) are the left (right) deoxyriboses. 
For the HOMO regime, the eigenvectors are depicted in Fig.~\ref{fig:V_N=5_HOMO}, in descending eigenenergy order. Eigenstate index 1 (15) corresponds to the highest (lowest) eigenenergy. 
Initially, in the \textit{upper panel}, we take $t_{\mathrm{S}} = t_{\mathrm{Sp}} = 0$. This results in 
tenfold degeneracy, as the ten deoxyriboses are completely isolated from the sequence of the five base pairs. 
The eigenenergies are
$-7.8268, -7.9000, -8.0000, -8.1000, -8.1732$ eV (four digits shown) and ten-fold degenerate exactly at $E_\mathrm{S} = -9$ eV. We observe that the first five eigenstates are localized at base pairs being of \textit{collective base-pair character} and the rest at exactly one deoxyribose each being of \textit{single deoxyribose character}. 

In the \textit{middle panel}, we take $t_{\mathrm{S}} = t_{\mathrm{Sp}} = 0.02$ eV. Now, the eigenenergies are $-7.8261, -7.8993, -7.9992,$ $-8.0991, -8.1722$ eV (four digits shown), five-fold degenerate (at our arithmetic accuracy) exactly at $E_\mathrm{S} = -9$ eV, and $-9.0007, -9.0007, -9.0008, -9.0009, -9.0010$ eV (four digits shown). We observe that now the first five eigenstates are still principally localized at base pairs being of \textit{collective base-pair character} and the rest principally at many deoxyriboses being of \textit{collective deoxyribose character}.
However, the five eigenstates with eigenenergies at $E_\mathrm{S} = -9$ eV have their two larger contributions at  the two deoxyriboses of the same monomer [(1,3), (4,6), (7,9), (10,12), (13,15)].
In the \textit{lower panel}, we take $t_{\mathrm{S}} = 0.02$ eV 
$\neq t_{\mathrm{Sp}} = 0.16$ eV. 

Now, the eigenenergies are
$-7.8050, -7.8769, -7.9746, -8.0720, -8.1429$ eV (four digits shown), five-fold degenerate (at our arithmetic accuracy) exactly at $E_\mathrm{S} = -9$ eV, and $-9.0218, -9.0231, -9.0254, -9.0280, -9.0303$ eV (four digits shown). We observe that still the first five eigenstates are mainly localized at base pairs being mainly of \textit{collective base-pair character}, the next five degenerate mainly at one deoxyribose each ($1,4,7,10,13$) being mainly of \textit{single deoxyribose character} and the rest five mainly at deoxyriboses $3,6,9,12,15$, being mainly of \textit{collective deoxyribose character}.

The values we take, $t_{\mathrm{S}} = 0.02$ eV, 
$t_{\mathrm{Sp}} = 0.16$ eV, serve to denote the relevant magnitude of interaction integrals between purine and backbone ($t_{\mathrm{S}} \sim 0.02$) and pyrimidine and backbone ($t_{\mathrm{Sp}} \sim 0.16$), as will be soon presented in another article, where the TB parameters were obtained by DFT with the method of Refs.~\cite{Baumeier:2010} and~\cite{Morphis:2023}.

\begin{figure}[h]
\centering
\includegraphics[width=0.45\textwidth]{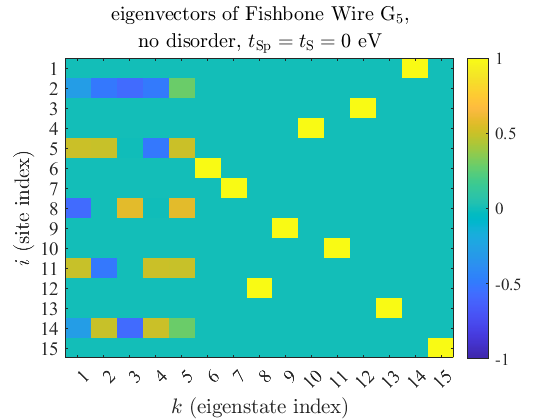} \\ \vspace{0.4cm} 
\includegraphics[width=0.45\textwidth]{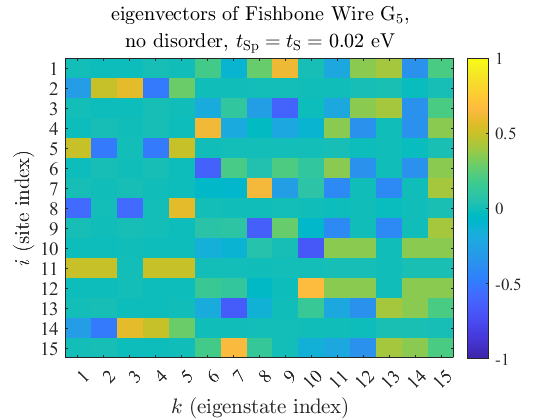} \\ \vspace{0.4cm}
\includegraphics[width=0.45\textwidth]{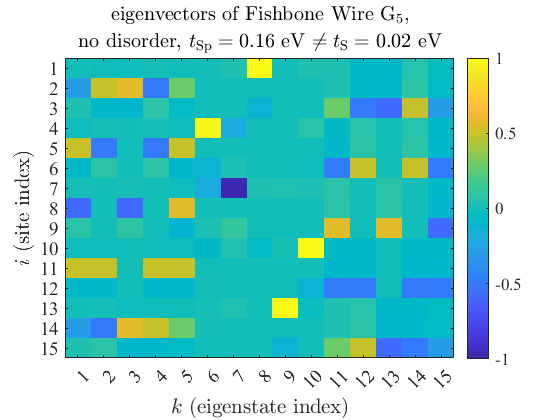}
\caption{TB Fishbone Wire Model. The eigenvectors of G$_5$ sequence. HOMO regime. Eigenstate index 1 (15) corresponds to the highest (lowest) eigenenergy. 
\textit{Upper panel:} $t_{\mathrm{S}} = t_{\mathrm{Sp}} = 0$.
\textit{Middle panel:} $t_{\mathrm{S}} = t_{\mathrm{Sp}} = 0.02$ eV. 
\textit{Lower panel:} $t_{\mathrm{S}} = 0.02$ eV 
$\neq t_{\mathrm{Sp}} = 0.16$ eV .}
\label{fig:V_N=5_HOMO}
\end{figure}

\begin{figure}
\centering
\includegraphics[width=0.45\textwidth]{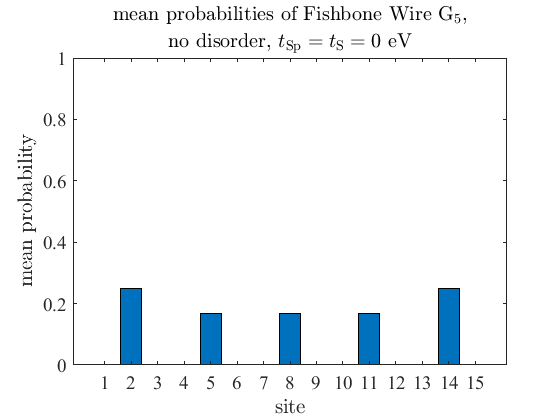} \\
\includegraphics[width=0.45\textwidth]{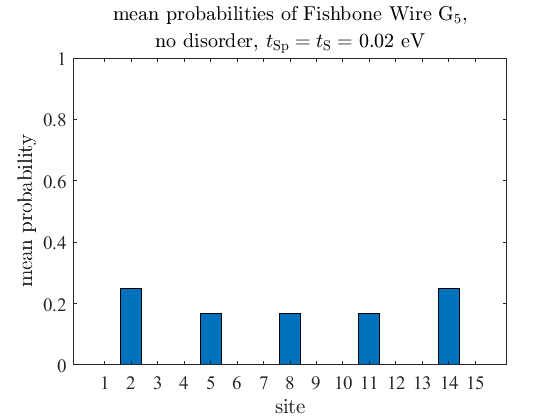} \\
\includegraphics[width=0.45\textwidth]{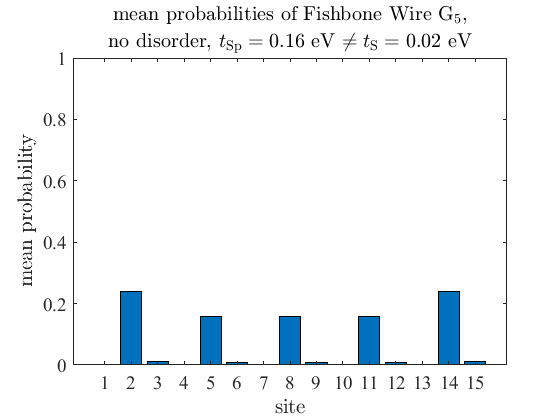} 
\caption{TB Fishbone Wire Model. G$_5$ sequence. Mean probabilities to find the hole at each site, after an initial oxidation at the second site (the first G-C base pair). \textit{Upper:} $t_{\mathrm{Sp}} = t_{\mathrm{S}} = 0$. This is exactly the TB Wire Model result \cite{LChMKThS:2015}: the end base pairs have mean probabilities 1/4 and the middle ones 1/6. \textit{Middle:} $t_{\mathrm{S}} = t_{\mathrm{Sp}} = 0.02$ eV. Now, it is almost like the Wire TB model result \cite{LChMKThS:2015}. \textit{Lower:} $ t_{\mathrm{S}} = 0.02$ eV $\neq t_{\mathrm{Sp}} = 0.16$ eV  leads to somehow larger mean probabilities at deoxyriboses connected to pyrimidines.}
\label{fig:MeanProbG5}
\end{figure}

The relevant mean over time probabilities to find the hole at each site, after an initial oxidation at the second site of G$_5$ (the first G-C base pair) are shown in Fig.~\ref{fig:MeanProbG5}. The basic message is that the hole predominantly remains at base pairs.
In the \textit{upper panel}, where $t_{\mathrm{Sp}} = t_{\mathrm{S}} = 0$, we obtain exactly the TB Wire Model result \cite{LChMKThS:2015}: the end base pairs have mean probabilities 1/4 and the middle ones 1/6. In the \textit{middle panel}, $t_{\mathrm{S}} = t_{\mathrm{Sp}} = 0.02$ eV, which leads to an almost TB Wire Model result \cite{LChMKThS:2015}; the mean probabilities are $\approx$ $0.0002, 0.2496, 0.0002$ for the 1st monomer, $0.0001, 0.1664, 0.0001$ for the 2nd monomer, $0.0001, 0.1664, 0.0001$ for the 3rd monomer, $0.0001, 0.1664, 0.0001$ for the 4th monomer, $0.0002, 0.2496, 0.0002$ for the 5th monomer. In the \textit{lower panel}, $ t_{\mathrm{S}} = 0.02$ eV $\neq t_{\mathrm{Sp}} = 0.16$ eV, which leads to somehow larger mean probabilities at deoxyriboses connected to pyrimidines:  $0.0002, 0.2380, 0.0118$ for the 1st monomer, $0.0001, 0.1585, 0.0080$ for the 2nd monomer, $0.0001, 0.1586, 0.0079$ for the 3rd monomer, $0.0001, 0.1585, 0.0080$ for the 4th monomer, $0.0002, 0.2381, 0.0118$ for the 5th monomer.

Let us now include disorder.
In Ref.~\cite{MMLSF:2021} Molecular Dynamics (MD) was used to estimate distortions and DFT and TB to evaluate their influence on the modification of on-site energies and interaction integrals. 
The base pair on-site energies are of the order of some eV (experimental ionization energies in the gas phase are around 8 eV) and their maximum variation is around 0.2 eV \cite{MMLSF:2021}. The interaction integrals present larger variations sometimes above 50\% \cite{MMLSF:2021}.
With this in mind, we allow for 5\% disorder at on-site energies, 50\% disorder at interaction integrals.
We run the random number generator 30 times to obtain each disordered series.
As a first choice, we keep the original (undisordered) interaction integral between base pairs and deoxyriboses $t_\mathrm{S} = t_\mathrm{S'} = 0.02$ eV on either side of the base pairs. As a second choice, we let 
$t_{\mathrm{S}} = 0.02$ eV $<$ $t_{\mathrm{S'}} = 0.16$ eV.
In Fig.~\ref{fig:PR+disorder-AllSources}, we show the effect of disorder, emanating from all possible sources, on participation ratio (PR). 
In Figs.~\ref{fig:PR+disorder-SeparateSources-tStS}- 
\ref{fig:PR+disorder-SeparateSources-tStSp}, in Appendix\ref{AppendixB}, we show the
effect of disorder on PR, from various separate sources. 
\begin{figure}[h]
\centering
\includegraphics[width=0.6\textwidth]{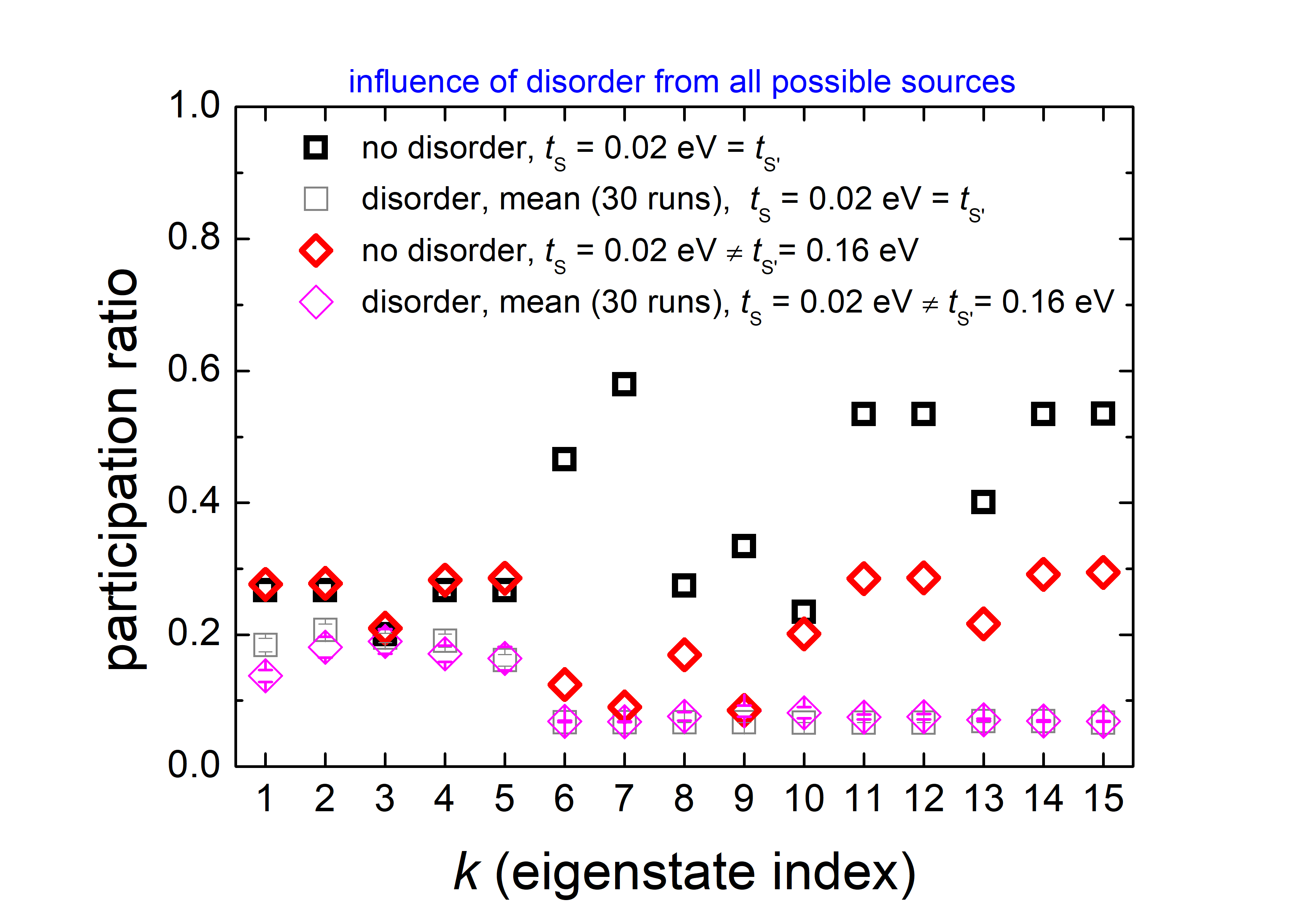} 
\caption{TB Fishbone Wire Model. G$_5$ sequence. Impact of disorder emanating from all possible sources on Participation Ratio. 
\textit{Squares:} $t_\mathrm{S} = t_\mathrm{S'} = 0.02$ eV; black bold squares: no disorder, 
gray squares: disorder.
\textit{Rhombuses:}
$t_{\mathrm{S}} = 0.02$ eV $<$ $t_{\mathrm{S'}} = 0.16$ eV; red bold rhombuses: no disorder,
magenda rhombuses: disorder.
Error bars are also shown.}
\label{fig:PR+disorder-AllSources}
\end{figure}

For the G$_5$ sequence and the TB Fishbone Wire Model,
Fig.~\ref{fig:ALLEigenenergies} displays the eigenenergies, 
Fig.~\ref{fig:ALLMeanProb} shows the mean probabilities at each site, and
Fig.~\ref{fig:kALL} presents the mean transfer rates from site 2 (first base pair), 
where initial oxidation takes place, to all other sites. In Fig.~\ref{fig:kALL},  
site 2 is shown with value 1; this is just a symbol to express that the hole was originally placed there. 
In the disordered cases, disorder from all possible sources has been taken into account and we display 
the mean over 30 runs with different random number generations. The possible sources of disorder are: the on-site energies of backbone sites (considered here simply as deoxyriboses), the on-site energies of base pairs, and the interaction integrals between base pairs as well as between base pairs and deoxyriboses. 
Mean value errors are also shown, but they are usually small on that scales.
In the \textit{upper panels},  $t_\mathrm{S} = t_\mathrm{S'} = 0.02$ eV, 
in the \textit{lower panels}, $t_{\mathrm{S}} = 0.02$ eV $<$ $t_{\mathrm{S'}} = 0.16$ eV.
Details on the influence of separate sources of disorder can be found in Ref.~\cite{Banev:2025}.

\begin{figure}
\centering
\hspace{-1cm}
\includegraphics[width=0.48\textwidth]{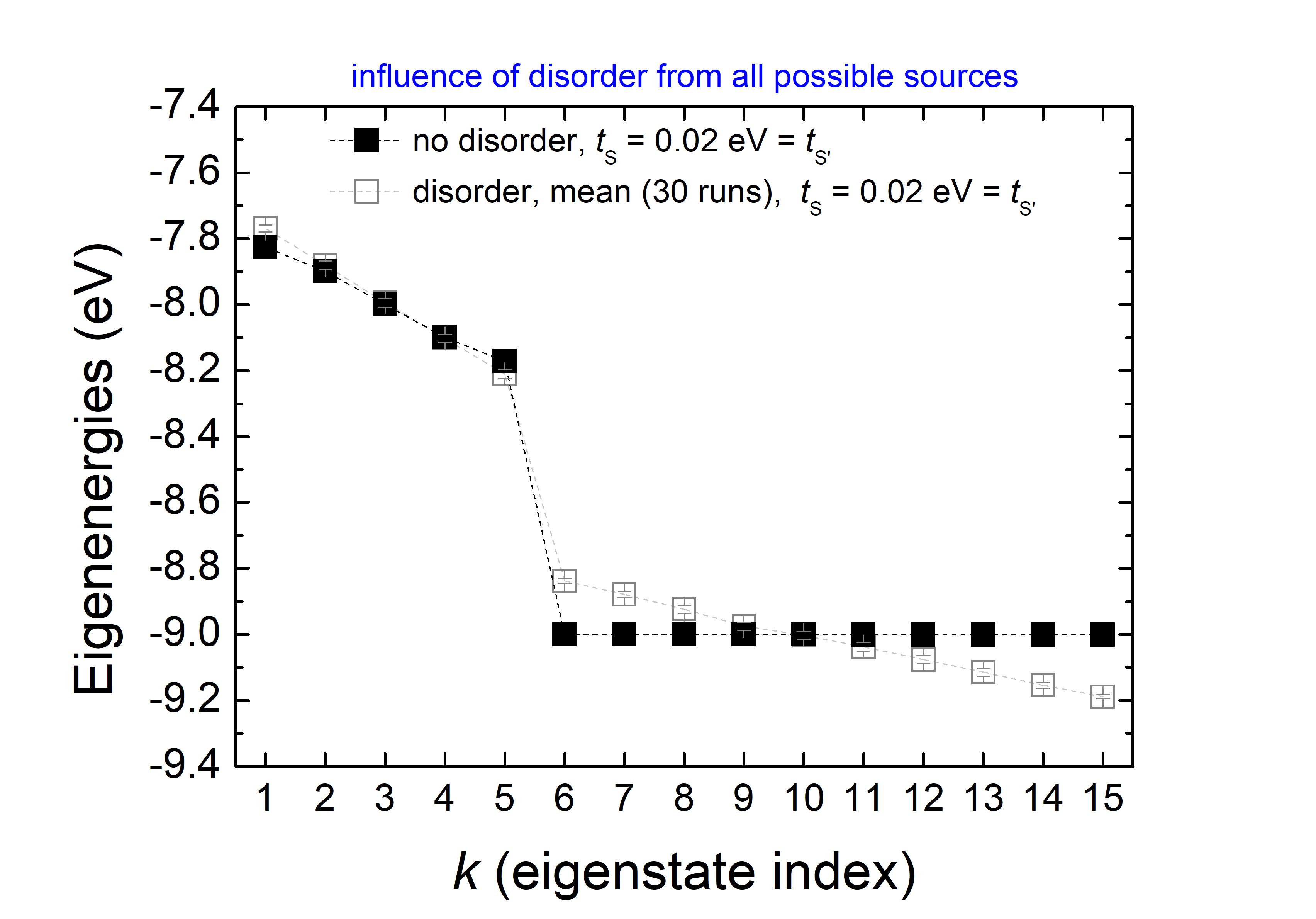}
\hspace{-1cm}
\includegraphics[width=0.48\textwidth]{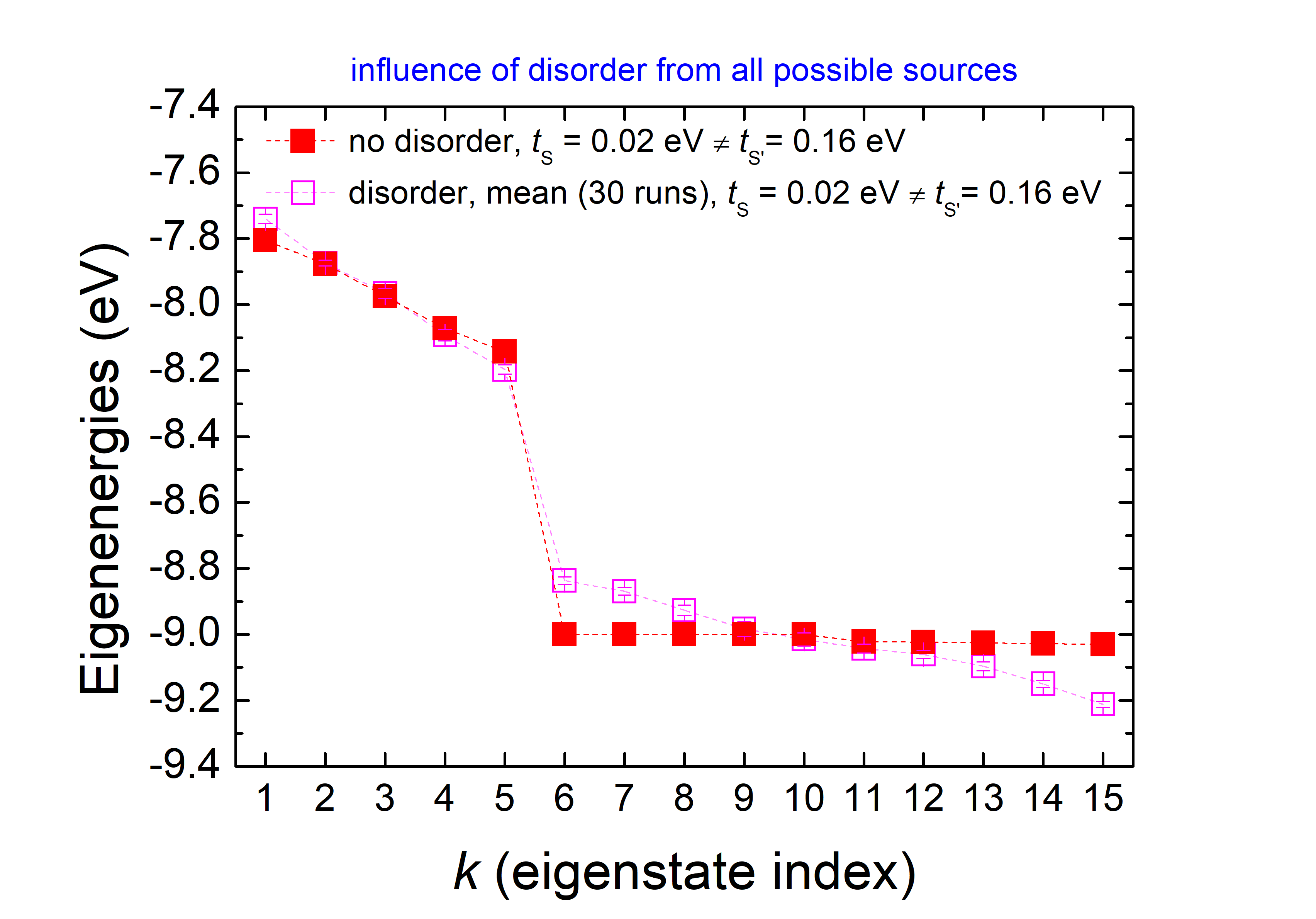}
\caption{TB Fishbone Wire Model. Eigenenergies of the G$_5$ sequence. 
\textit{Left.}  $t_\mathrm{S} = t_\mathrm{S'} = 0.02$ eV; black bold squares: no disorder, gray squares: disorder.
\textit{Right.} $t_{\mathrm{S}} = 0.02$ eV $<$ $t_{\mathrm{S'}} = 0.16$ eV; red bold rhombuses: no disorder, magenda rhombuses: disorder. Disorder from all possible sources has been taken into account. Error bars are also shown.}
\label{fig:ALLEigenenergies}
\end{figure}
\begin{figure}
\centering
\hspace{-1cm}
\includegraphics[width=0.48\textwidth]{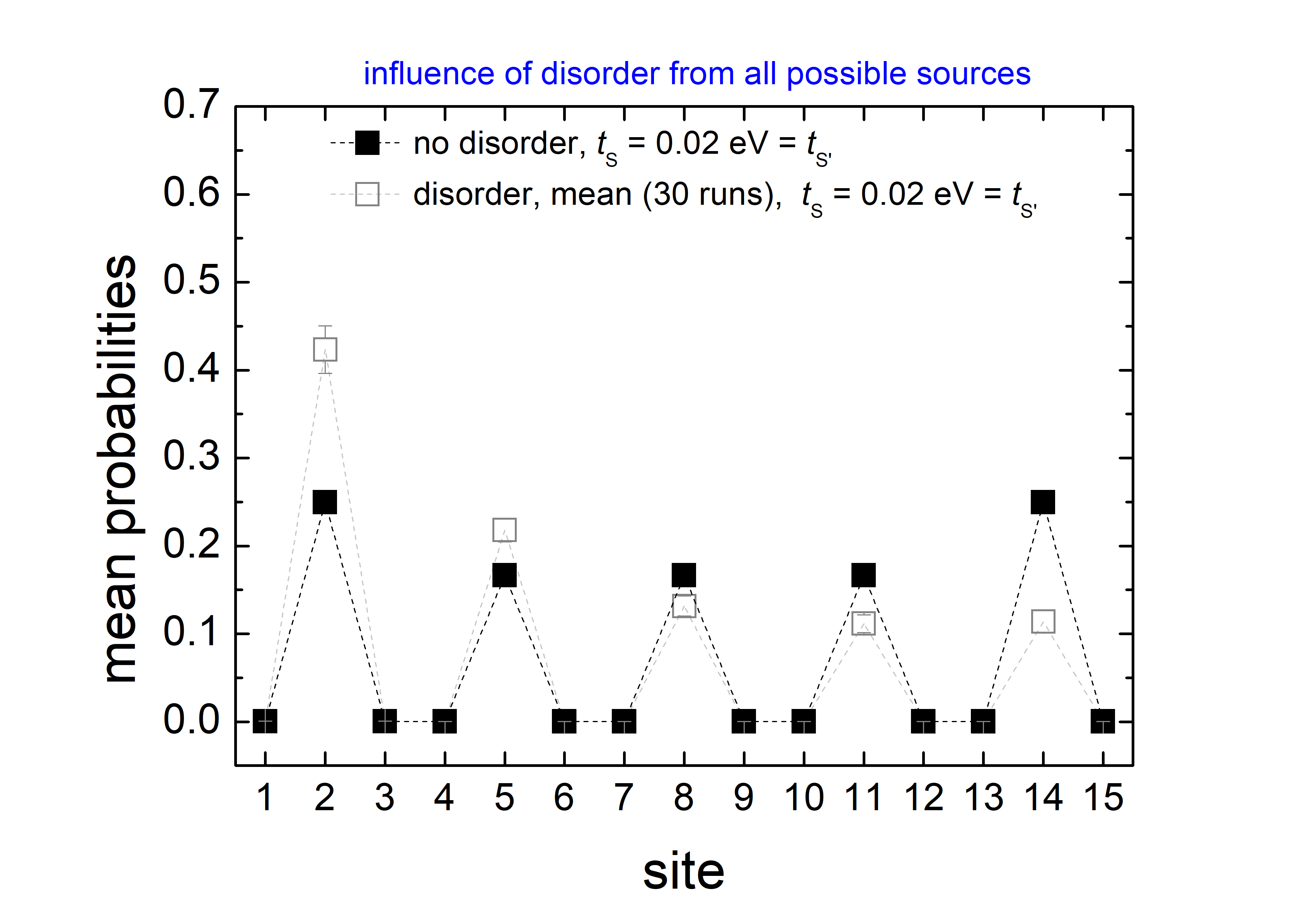}
\hspace{-1cm}
\includegraphics[width=0.48\textwidth]{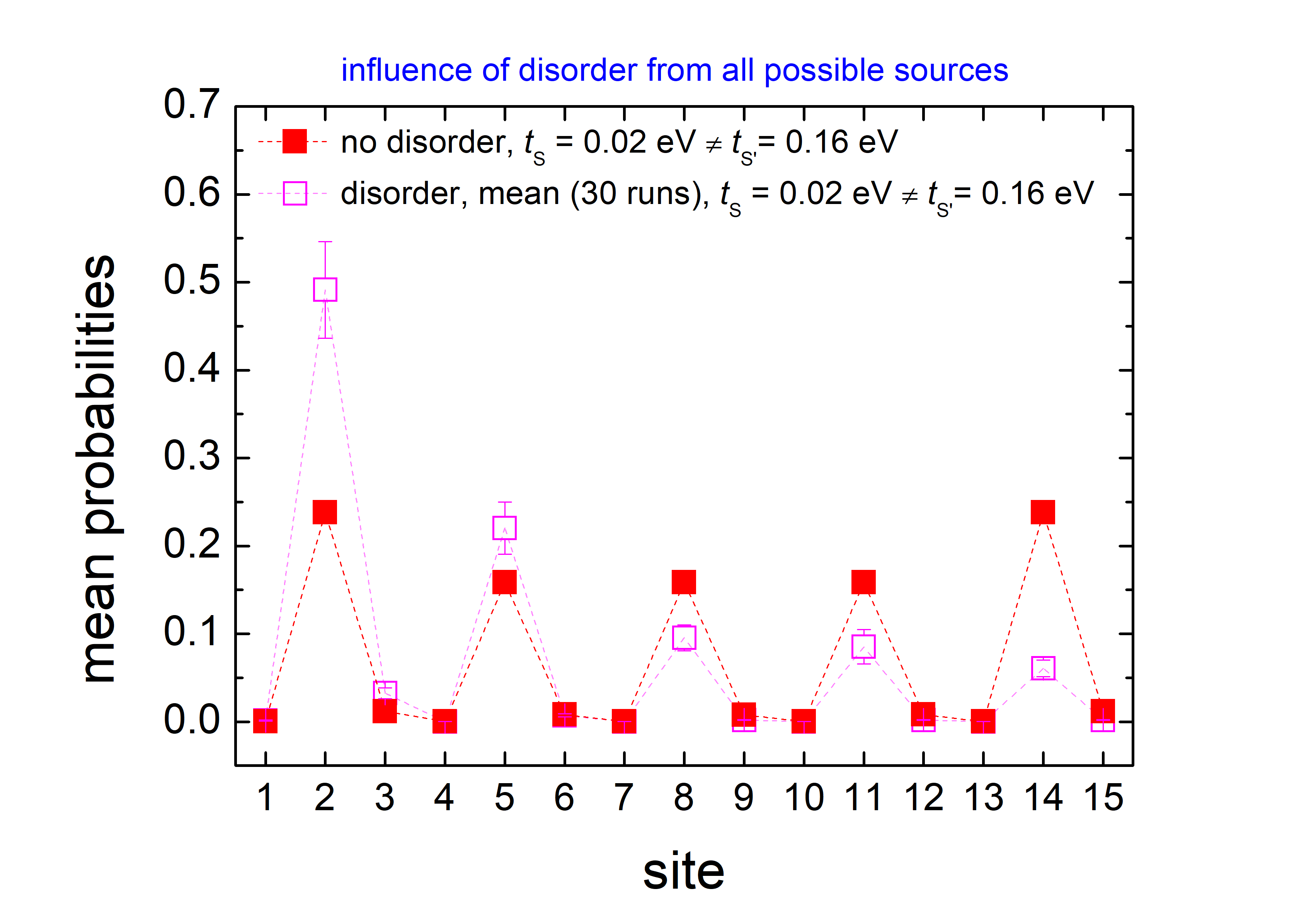}
\caption{TB Fishbone Wire Model. Mean probabilities at each site for G$_5$ sequence. 
\textit{Left.}  $t_\mathrm{S} = t_\mathrm{S'} = 0.02$ eV; black bold squares: no disorder, gray squares: disorder.
\textit{Right.} $t_{\mathrm{S}} = 0.02$ eV $<$ $t_{\mathrm{S'}} = 0.16$ eV; red bold rhombuses: no disorder, magenda rhombuses: disorder. Disorder from all possible sources has been taken into account. Error bars are also shown.}
\label{fig:ALLMeanProb}
\end{figure}
\begin{figure}
\centering
\hspace{-1cm}
\includegraphics[width=0.48\textwidth]{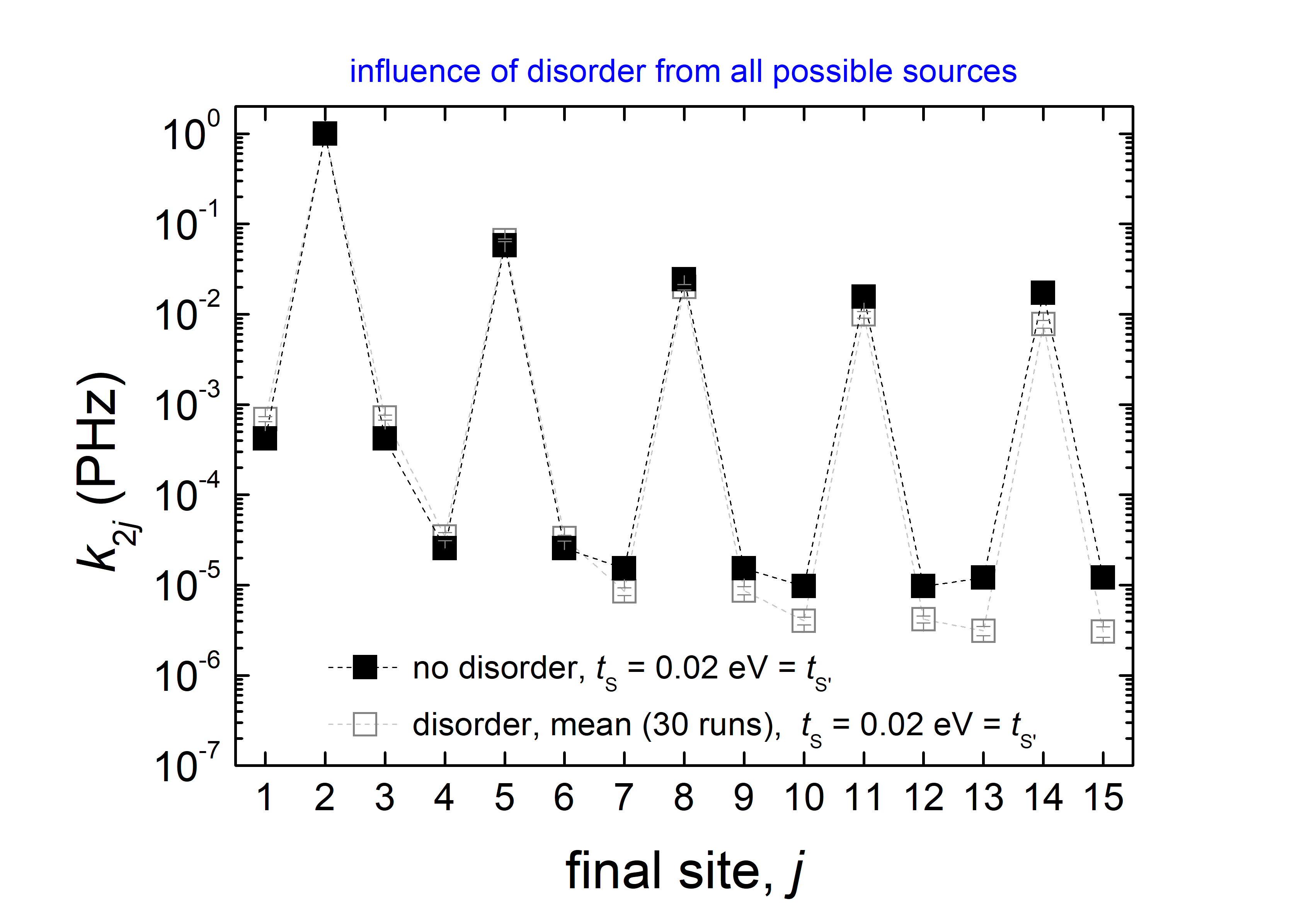}
\hspace{-1cm}
\includegraphics[width=0.48\textwidth]{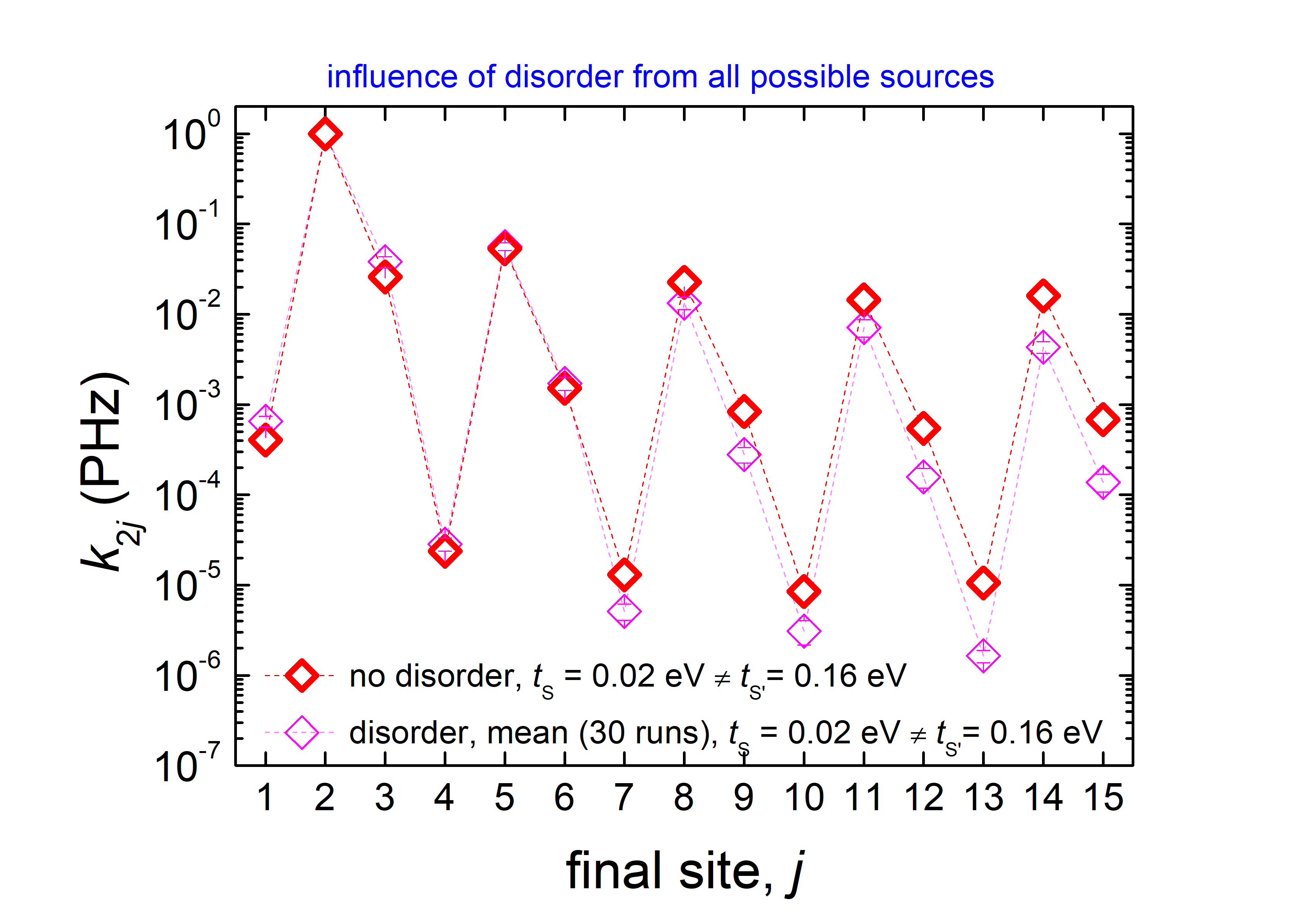}
\caption{TB Fishbone Wire Model. 
Mean transfer rates from site 2 (initial oxidation at the first base pair) to all other sites for the G$_5$ sequence. 
\textit{Left.}  $t_\mathrm{S} = t_\mathrm{S'} = 0.02$ eV; black bold squares: no disorder, gray squares: disorder.
\textit{Right.} $t_{\mathrm{S}} = 0.02$ eV $<$ $t_{\mathrm{S'}} = 0.16$ eV; red bold rhombuses: no disorder, magenda rhombuses: disorder. Site 2 is shown with value 1, this is just a symbol to express that the hole was originally placed there. Disorder from all possible sources has been taken into account. Error bars are also shown.}
\label{fig:kALL}
\end{figure}

\textit{Conclusion:} Disorder from all possible sources has been taken into account. Disorder leads to severe modifications of participation ratios, i.e., it leads to increase of localization. Relevant modifications also occur on eigenenergies, mean probabilities at each site, and transfer rates.

\section{\label{sec:ConclusionPerspectives} Conclusion}
We described the impact of transition mutations and disorder on charge transfer along B-DNA sequences. 
Since homopolymers are victores ludorum of charge transfer~\cite{LVBMS:2018, MLTS:2019}, we used homopolymers as starting, ideal, flawless sequences. Then we disturbed these ideal sequences, introducing either transition mutations or disorder. In this work we excluded the possibility of charge transfer via the backbone that will hopefully be addressed soon in another work. We employed the Tight Binding (TB) Wire model to study the influence of transition mutations and the TB Fishbone Wire model to evaluate the influence of  disorder emanating either from the $\pi$ path or from the backbone. 

We calculated the HOMO and LUMO eigenenergies and eigenvectors, the participation ratio, the time-dependent probability to find the carrier at each site, the mean over time probability at each site and the mean transfer rate from site to site.

We described how mutations increase localization as seen from participation ratio and how they impede charge transfer as seen from mean probability at each site and from the transfer rates from site to site.

We considered disorder from all possible sources: on-site energies of base pairs, 
on-site energies of backbone sites, 
interaction integrals between sites.
We described how disorder leads to modification of participation ratios, i.e., increase of localization. 
Moreover, due to disorder, modifications occur on eigenenergies, mean probabilities at each site, and transfer rates from site to site.

\vspace{-0.3cm}
\section*{\label{sec:AuthorContributions} Author Contributions}
Conceptualization, CS; methodology, CS; software, CS; validation, CS; formal analysis, AF, PB, CS; investigation, AF, PB, CS; resources, CS;
data curation, AF, PB, CS; writing – original draft preparation,
CS; writing – review and editing, CS; visualization, CS; supervision,
CS; project administration, CS; funding acquisition, CS.

\vspace{-0.3cm}

\section*{\label{sec:acknowledgments} Acknowledgments}
We thank the RISE worldwide Deutscher Akademischer Austauschdienst (DAAD) BSc student Amelie Schmohl from Technische Universit\"at Dresden, Germany, who made her practice from 4 Sep to 15 Oct 2023 in our Group, for checking once again the matlab code of the TB Wire Model and RISE worldwide DAAD BSc student Tina Ruckdeschel from Friedrich-Schiller-Universität Jena, Germany, who made her practice from 19 Aug to 11 Oct 2024 in our Group, for checking once again our calculations of participation ratio.

\clearpage

\vspace{-0.5cm}


\appendix

	
\section{TB Wire Model}
\label{AppendixA}
\setcounter{table}{0}  
\renewcommand{\thetable}{A\arabic{table}} 
\setcounter{figure}{0} 
\renewcommand{\thefigure}{A\arabic{figure}}
	
Introducing  G$\to$A transitions, important modifications of the TB parameters occur only in the HOMO regime, cf.~Tables~\ref{Table:OnsiteEnergies} and
\ref{Table:InteractionIntegrals}. We show the HOMO and LUMO regime eigenspectra in Fig.~\ref{fig:EigenSpectra}.
	
\begin{figure}[h]
\centering
\includegraphics[width=0.45\textwidth]{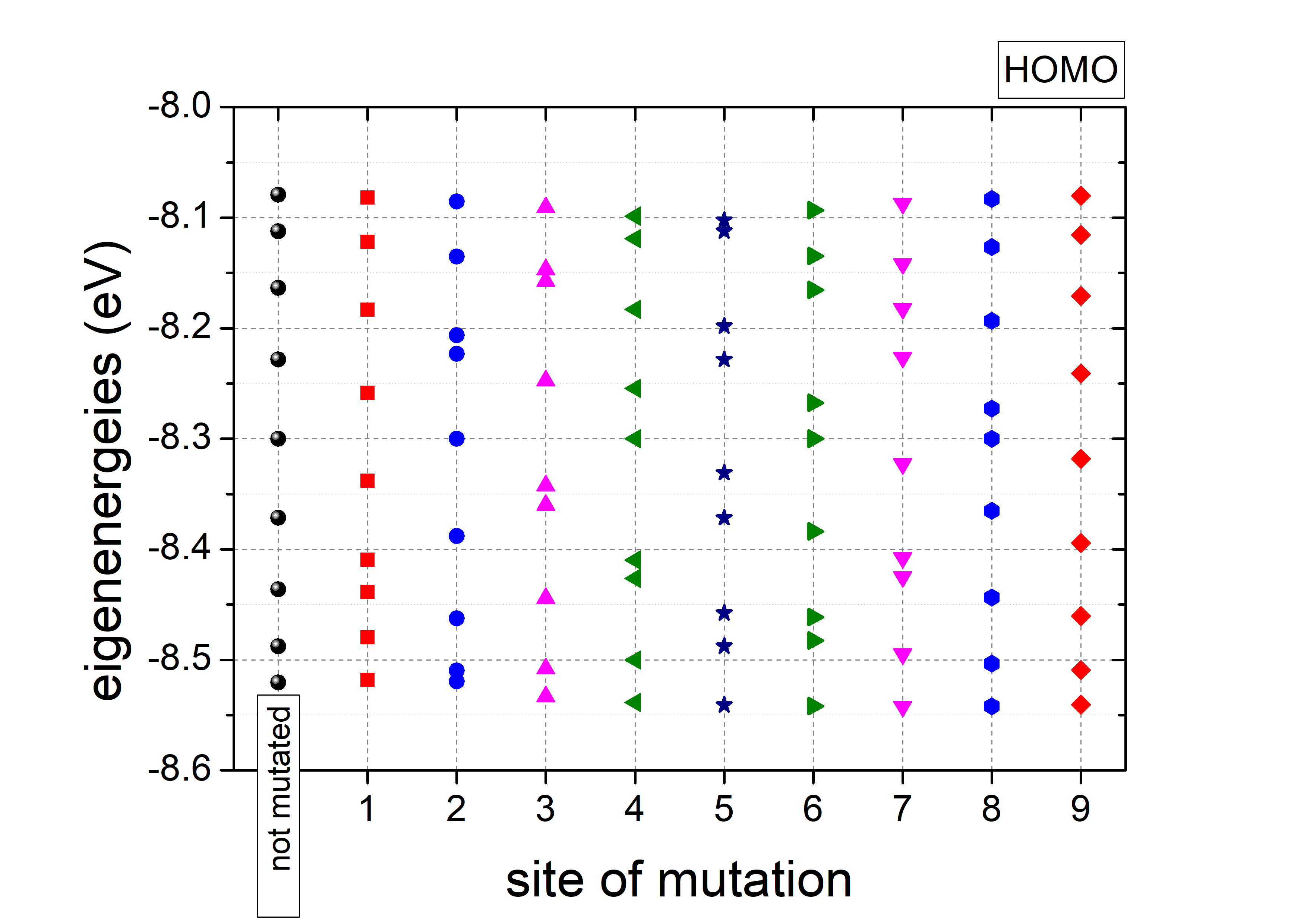}
\includegraphics[width=0.45\textwidth]{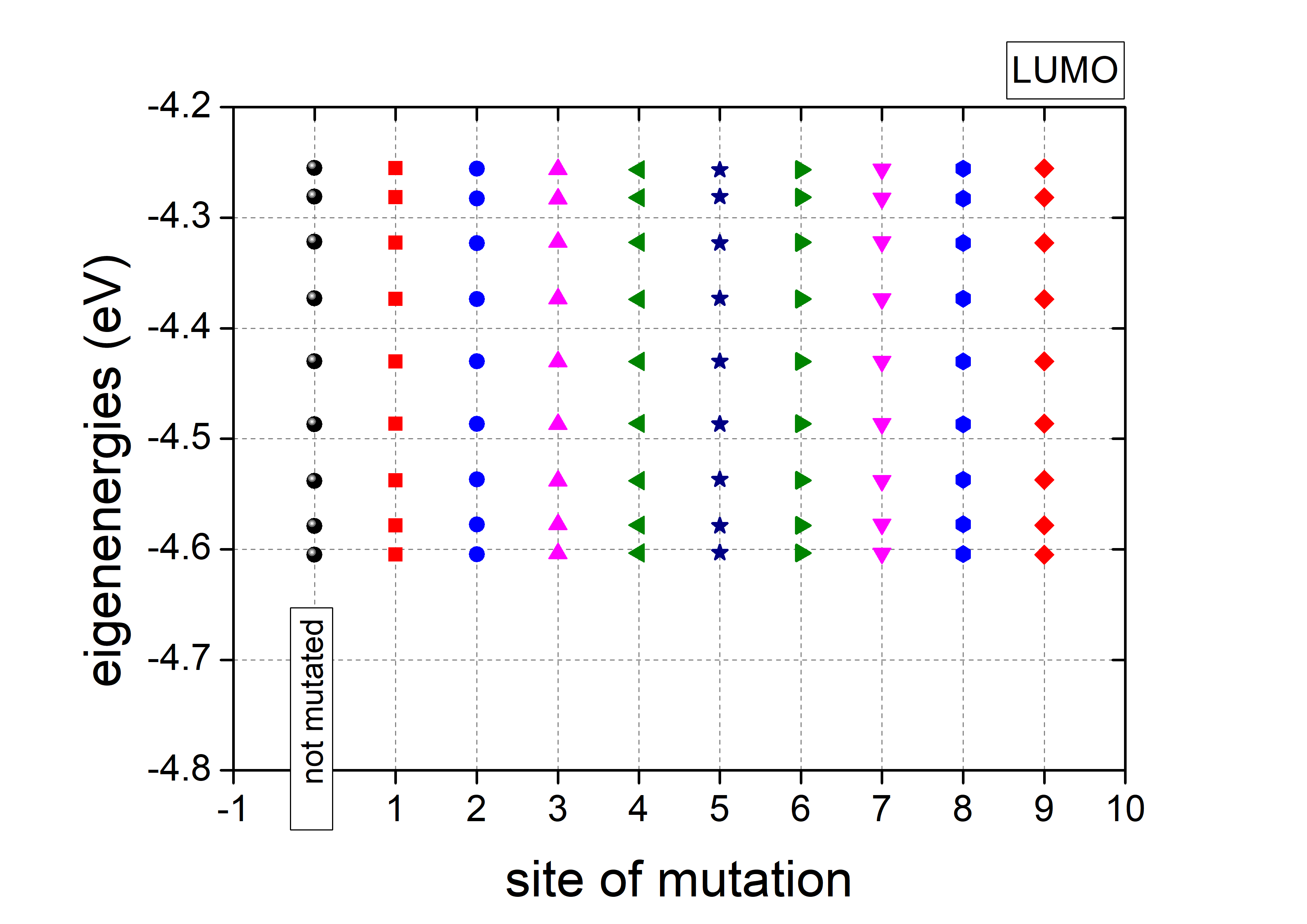} \\ 
\includegraphics[width=0.45\textwidth]{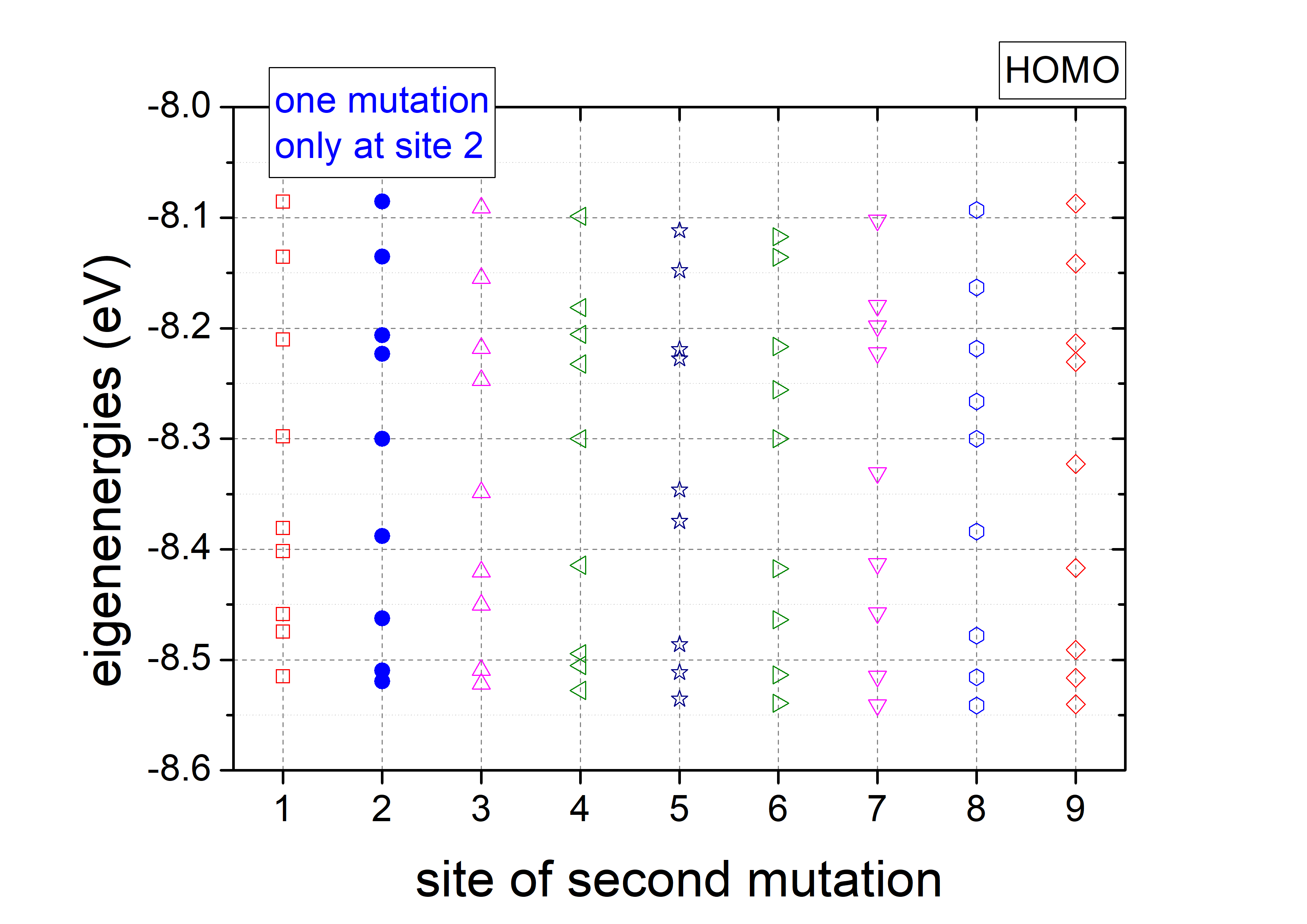}
\includegraphics[width=0.45\textwidth]{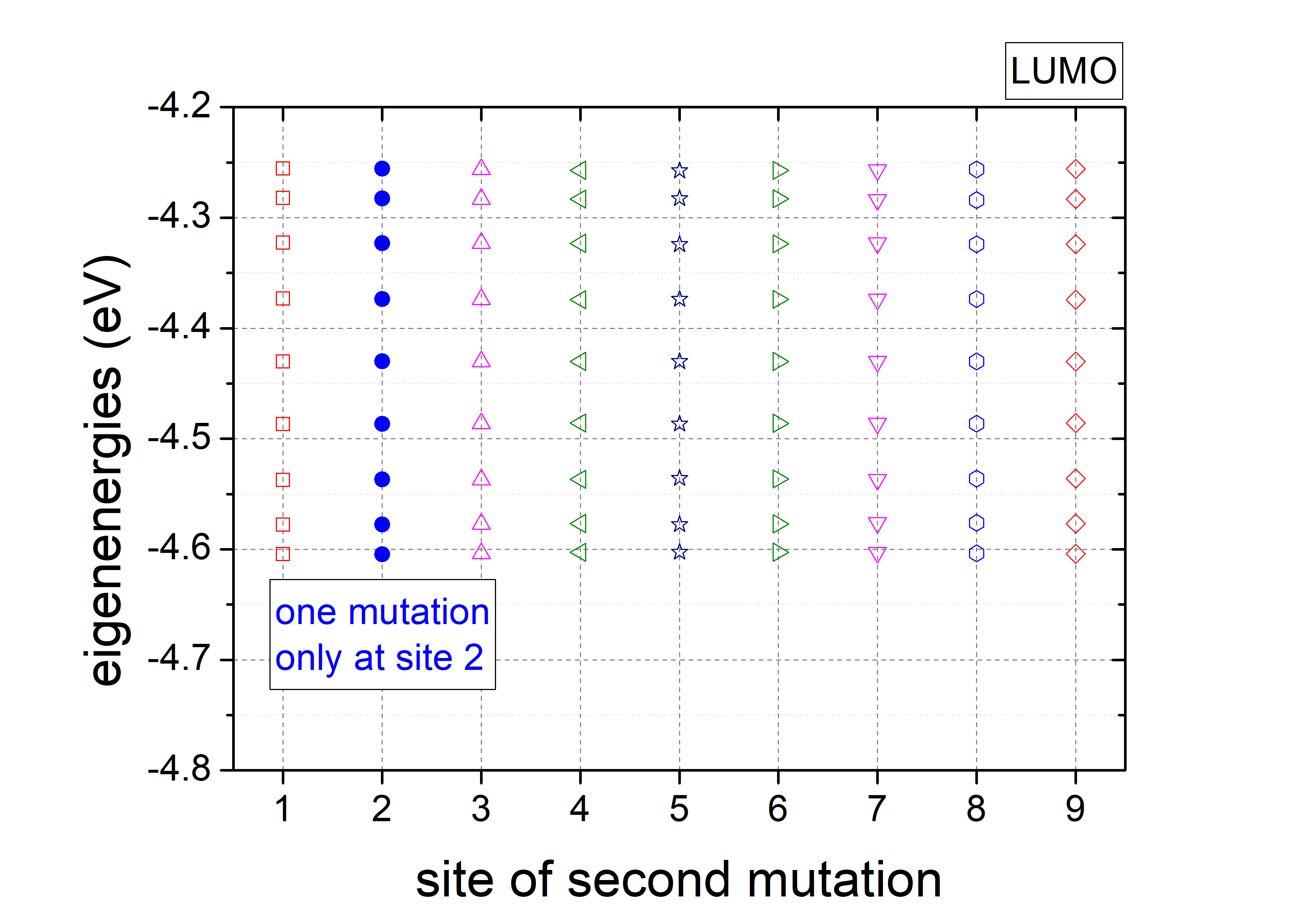} \\ 
\includegraphics[width=0.45\textwidth]{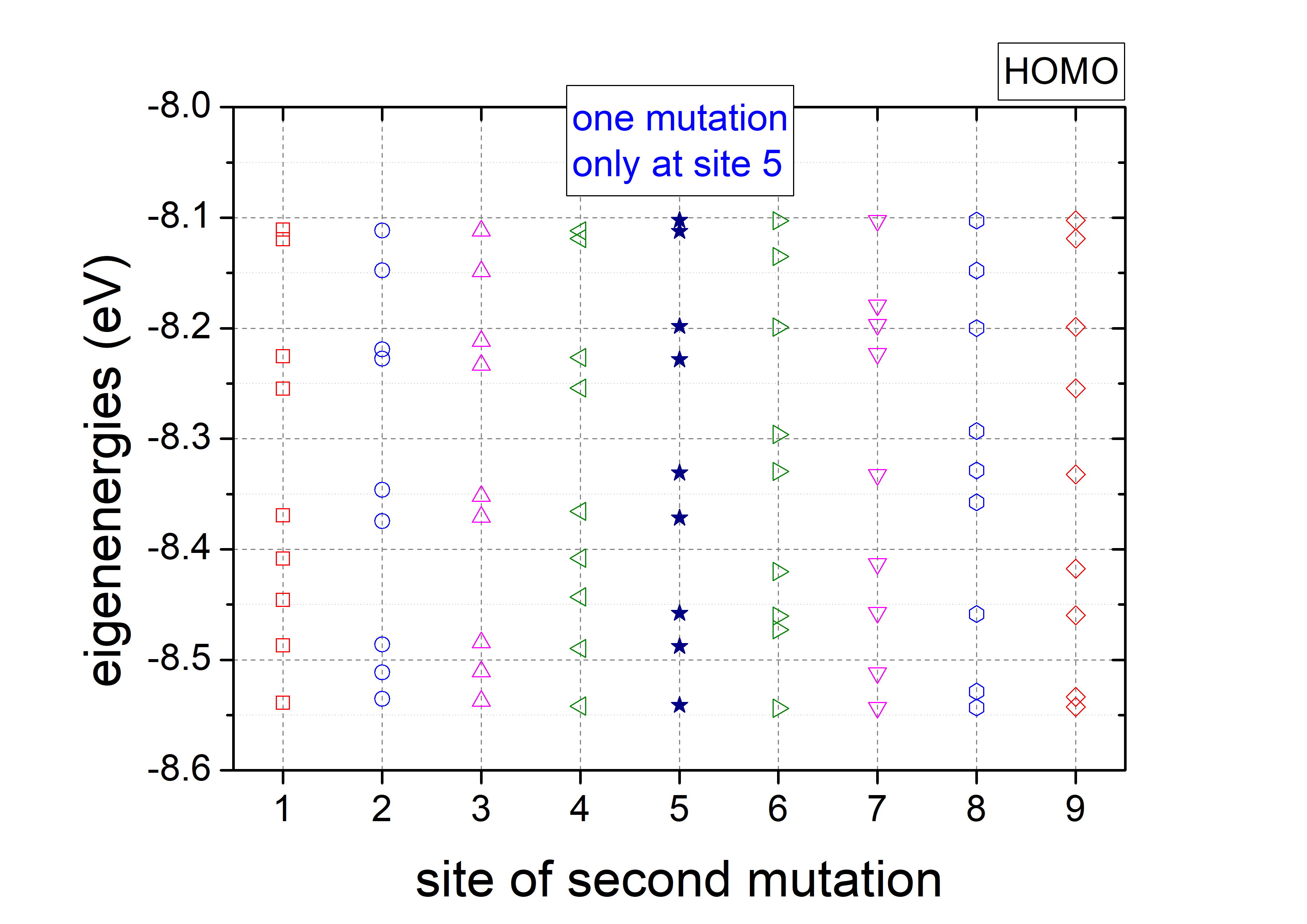}
\includegraphics[width=0.45\textwidth]{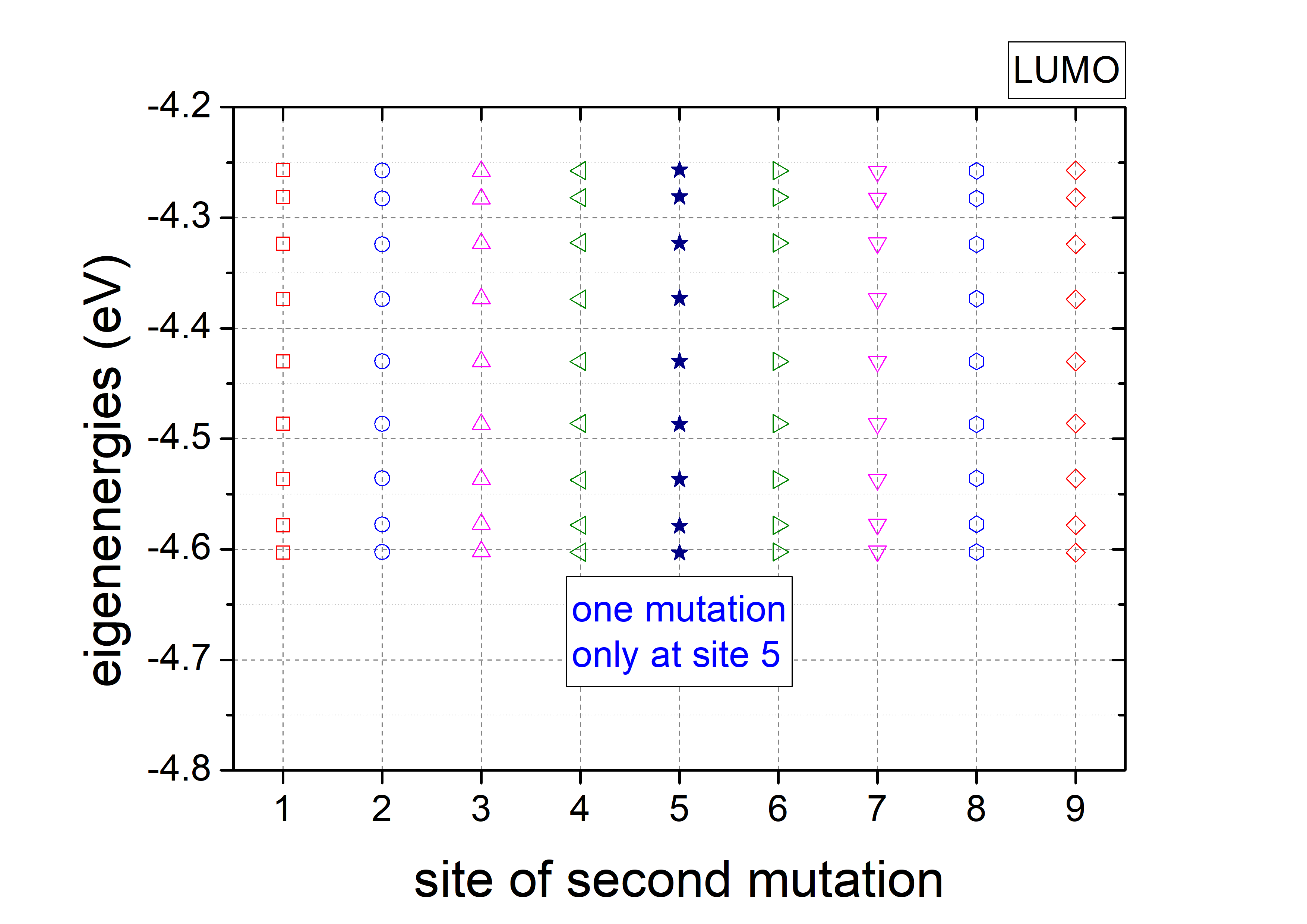}\caption{TB Wire Model. Eigenspectra modification due to one or two point mutations (G$\leftrightarrow$A transitions) at an initially ideal G$_9$ sequence, for the HOMO and LUMO regime, i.e., for oxidation and reduction, respectively.
First line: one mutation of varying position.
We also depict the not mutated case, G$_9$.
Second line: two mutations, one stable at site 2 and one of varying position. 
We also depict the case with only one mutation at site 2.
Third line: two mutations, one stable at site 5 and one of varying position. We also depict the case with only one mutation at site 5.}
\label{fig:EigenSpectra}
\end{figure}

\begin{figure}
\centering
\includegraphics[width=0.32\textwidth]{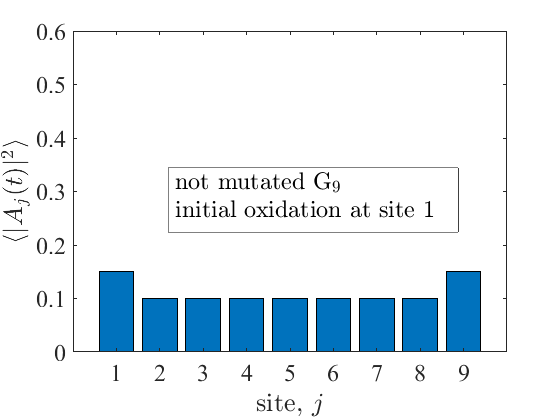}
\includegraphics[width=0.32\textwidth]{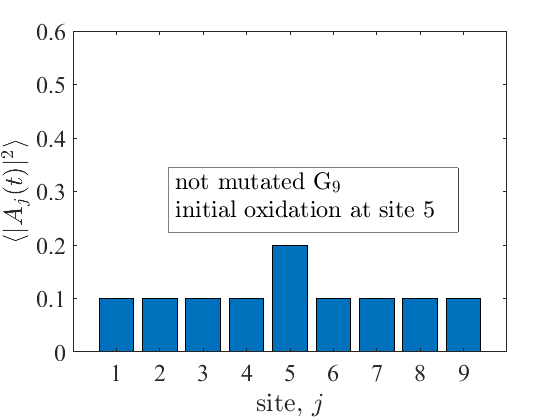}\caption{TB Wire Model. G$_9$ without mutation. The probabilities to find the hole at each site, after an initial oxidation at site 1 (left) and site 5 (right).}
\label{fig:MeanProbio1+io5}
\end{figure}
	
\begin{figure*}
\centering
\includegraphics[width=0.32\textwidth]{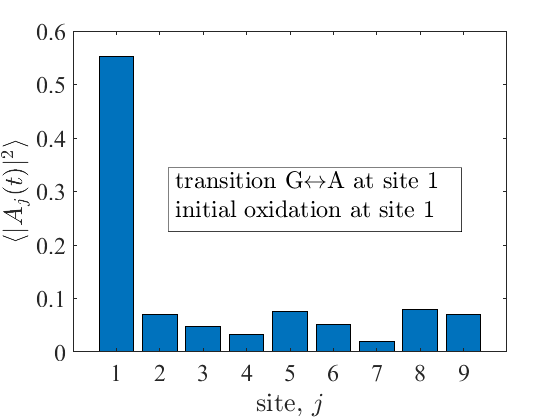}
\includegraphics[width=0.32\textwidth]{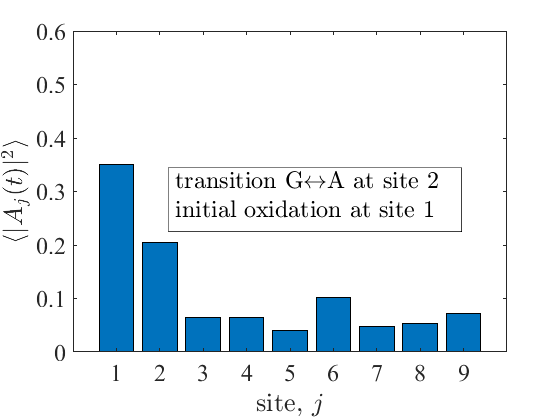}
\includegraphics[width=0.32\textwidth]{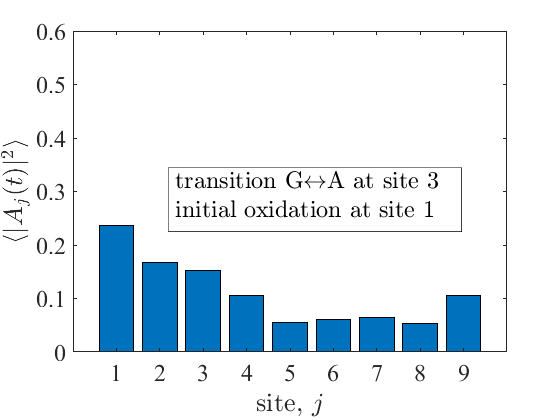}
\includegraphics[width=0.32\textwidth]{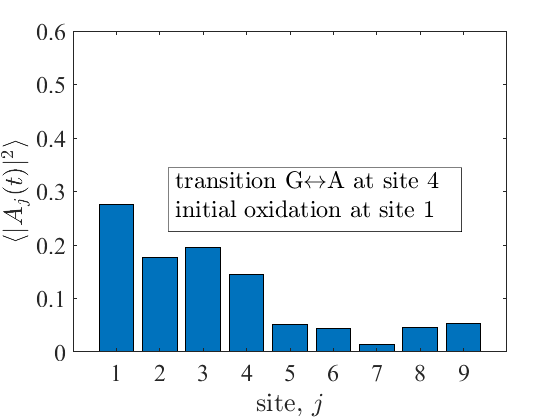}
\includegraphics[width=0.32\textwidth]{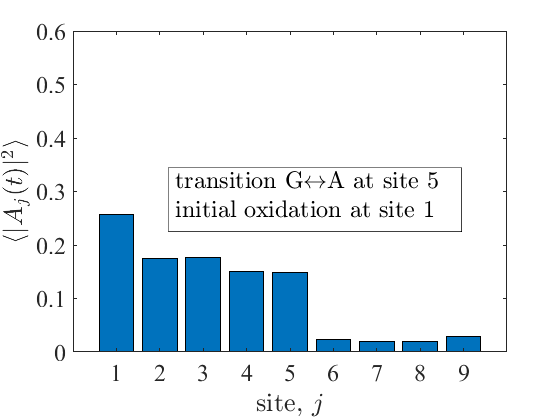}
\includegraphics[width=0.32\textwidth]{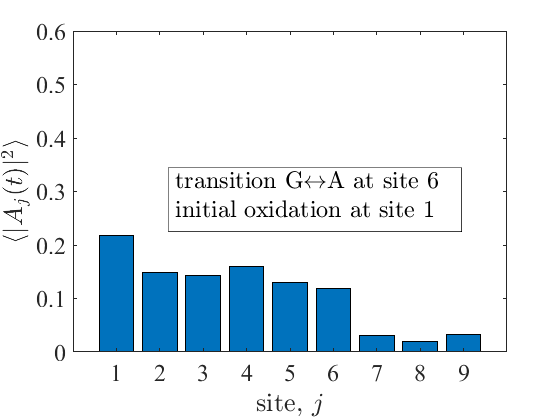}
\includegraphics[width=0.32\textwidth]{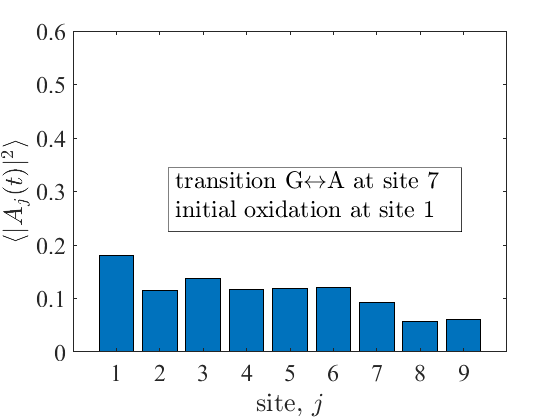}
\includegraphics[width=0.32\textwidth]{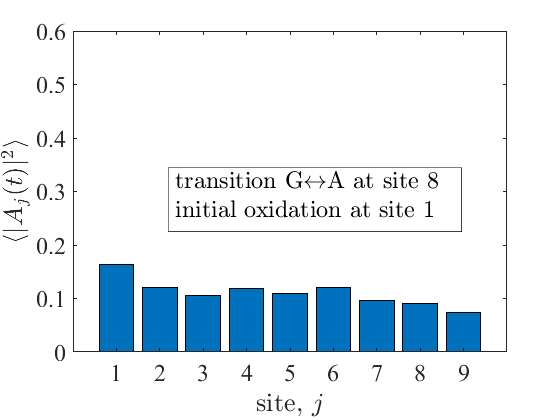}
\includegraphics[width=0.32\textwidth]{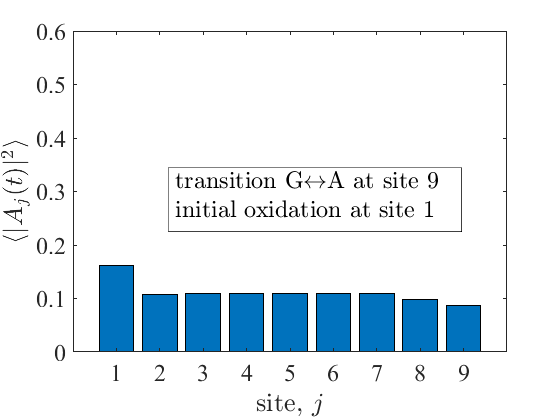}
\caption{TB Wire Model. G$_9$ and G$_9$ with one point mutation (transition G$\leftrightarrow$A) of varying position in the sequence. The probabilities to find the hole at each site, after an initial oxidation at site 1.}
\label{fig:MeanProbio1}
\end{figure*}
	
\begin{figure*}
\centering
\includegraphics[width=0.32\textwidth]{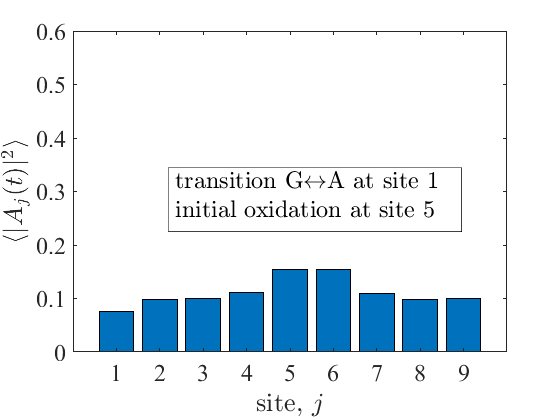}
\includegraphics[width=0.32\textwidth]{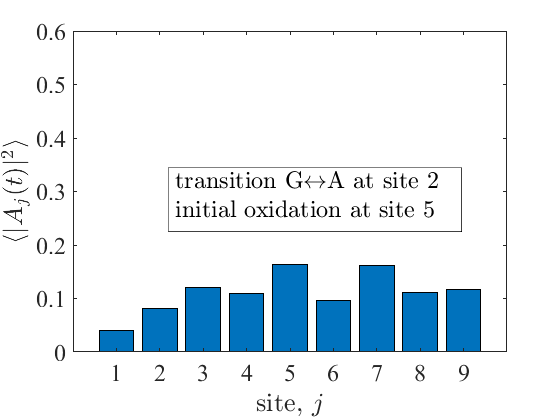}
\includegraphics[width=0.32\textwidth]{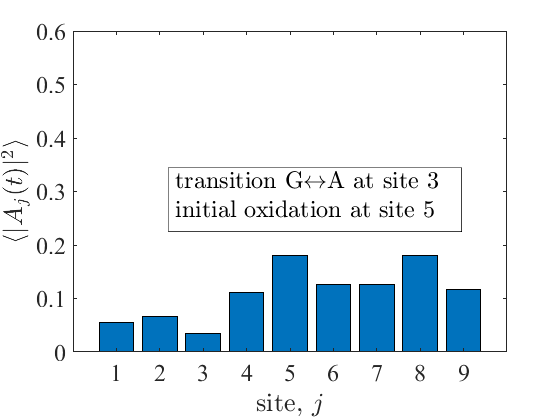}
\includegraphics[width=0.32\textwidth]{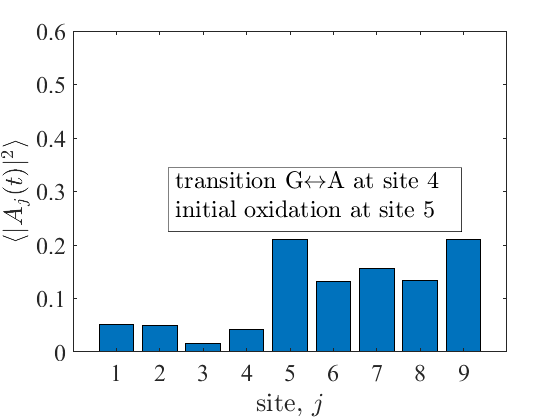}
\includegraphics[width=0.32\textwidth]{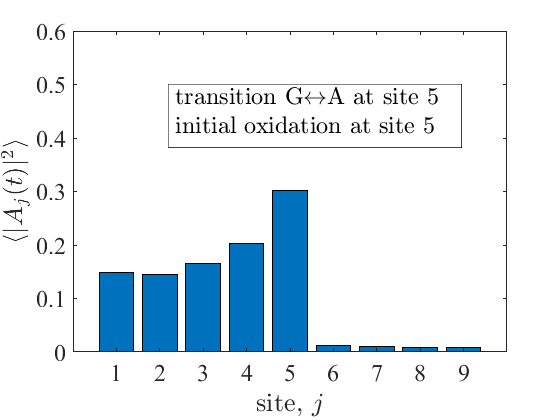}
\includegraphics[width=0.32\textwidth]{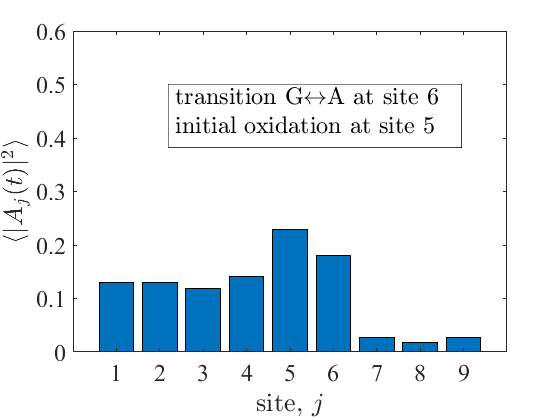}
\includegraphics[width=0.32\textwidth]{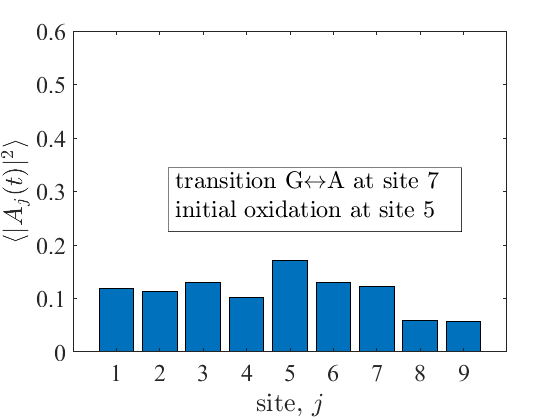}
\includegraphics[width=0.32\textwidth]{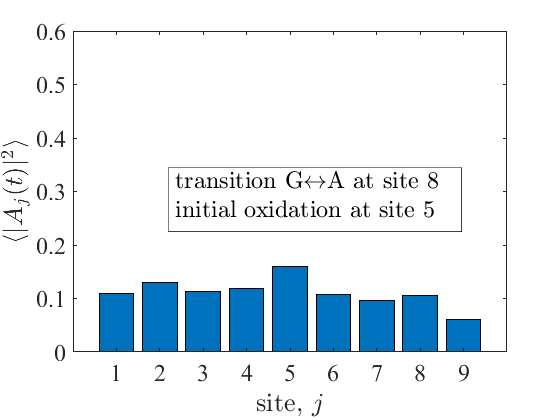}
\includegraphics[width=0.32\textwidth]{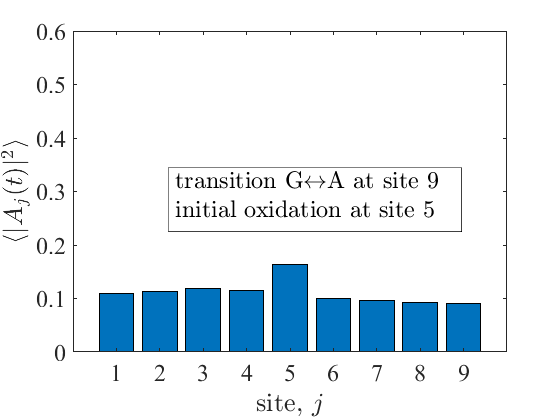}
\caption{TB Wire Model. G$_9$ and G$_9$ with one point mutation (transition G$\leftrightarrow$A) of varying position in the sequence. The probabilities to find the hole at each site, after an initial oxidation at site 5.}
\label{fig:MeanProbio5}
\end{figure*}

\clearpage
	
\section{Participation ratio: disorder from various sources}
\label{AppendixB}
\setcounter{table}{0}  
\renewcommand{\thetable}{B\arabic{table}} 
\setcounter{figure}{0} 
\renewcommand{\thefigure}{B\arabic{figure}}
	
In Figs.~\ref{fig:PR+disorder-SeparateSources-tStS}- 
\ref{fig:PR+disorder-SeparateSources-tStSp}, we show the
effect of disorder on PR, from various separate sources. 
We  ran the random number generator 30 times to obtain each disorder graph. Here we show the mean value and the mean error (usually small on that scale). We also tried all possible combinations of disorder sources, but we do not show these results, for brevity, as nothing additional was observed. Some details can be found in Ref.~\cite{Banev:2025}.

\begin{figure}[h]
\centering
\includegraphics[width=0.48\textwidth]{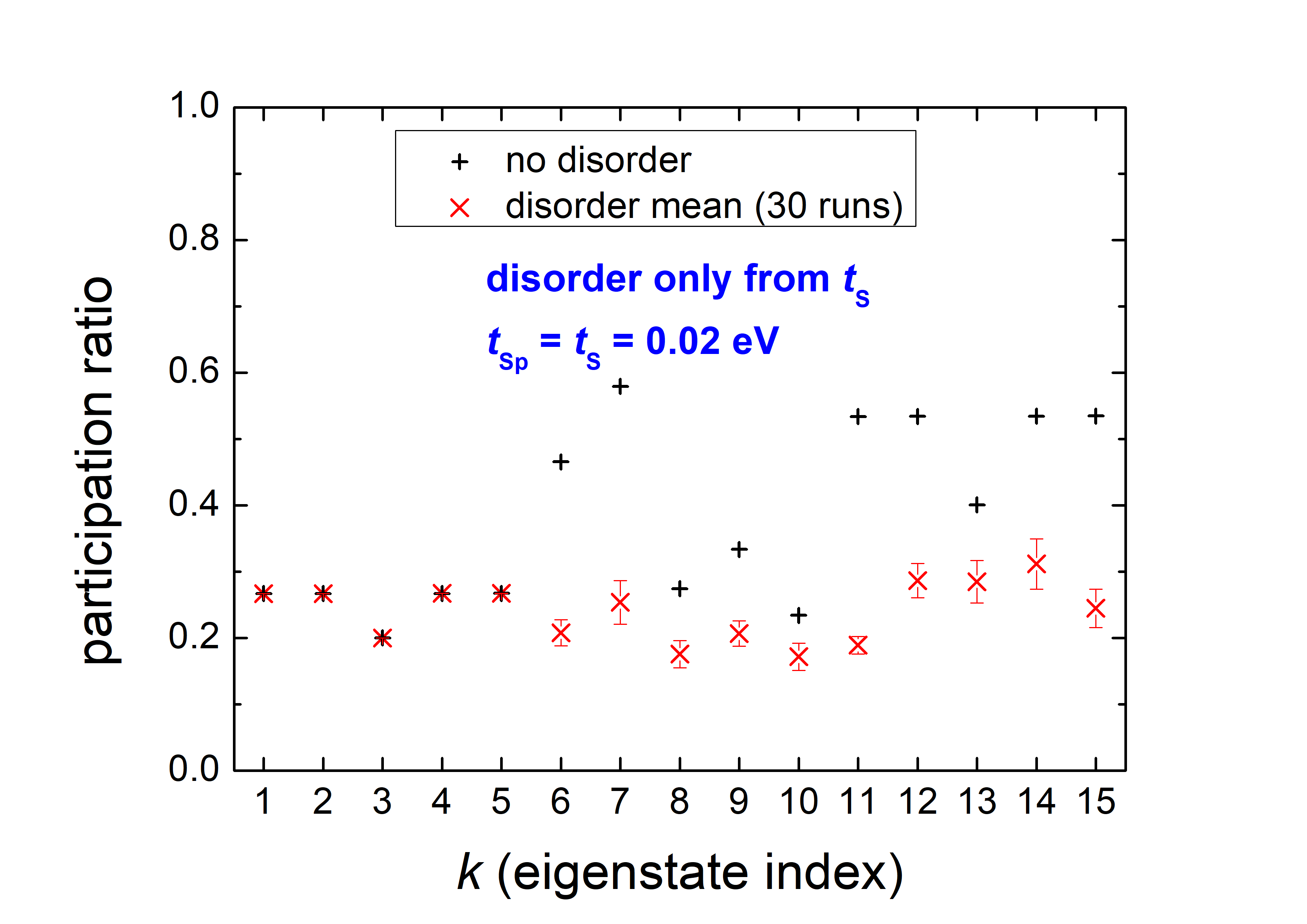} 
\includegraphics[width=0.48\textwidth]{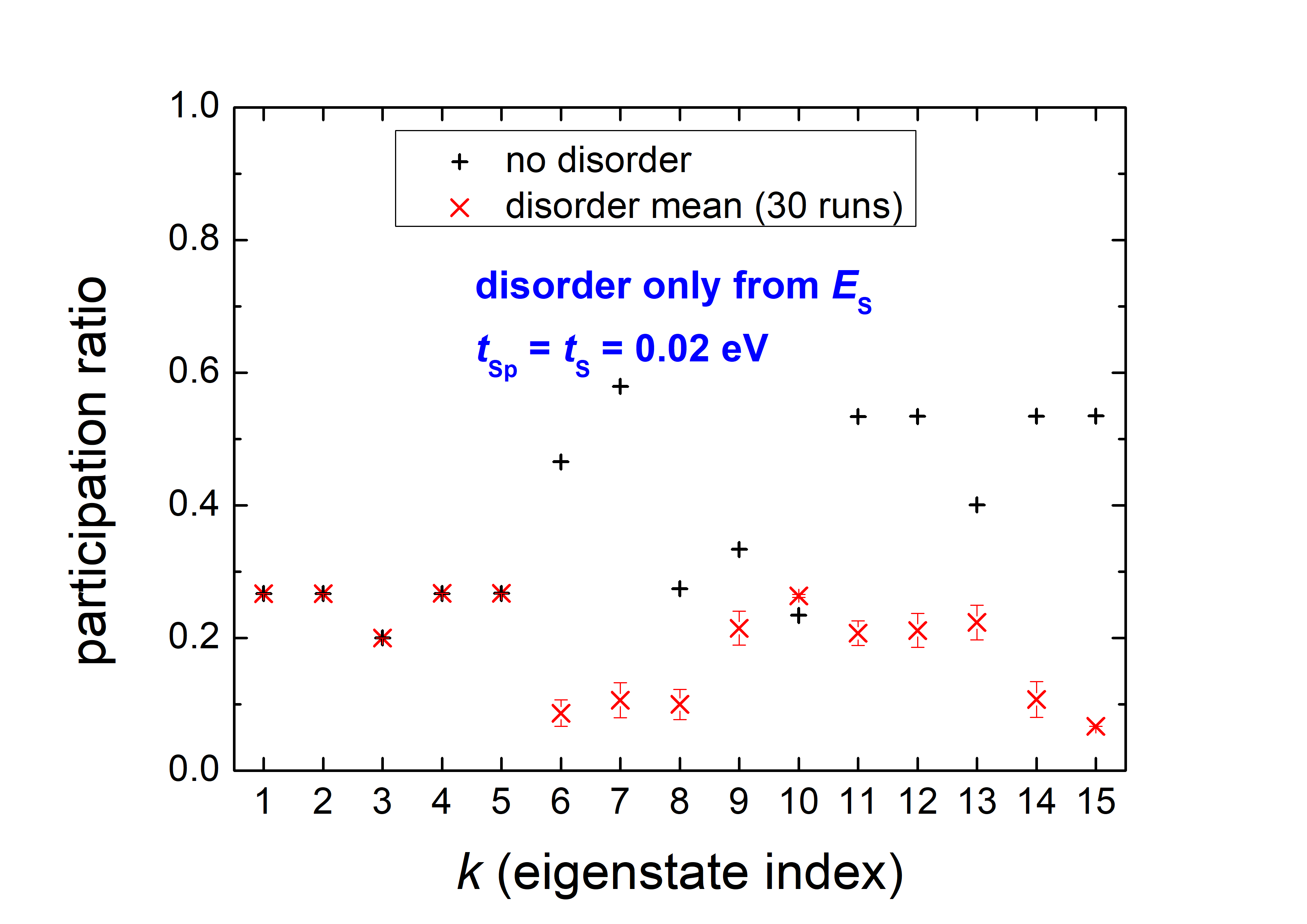} 
\includegraphics[width=0.48\textwidth]{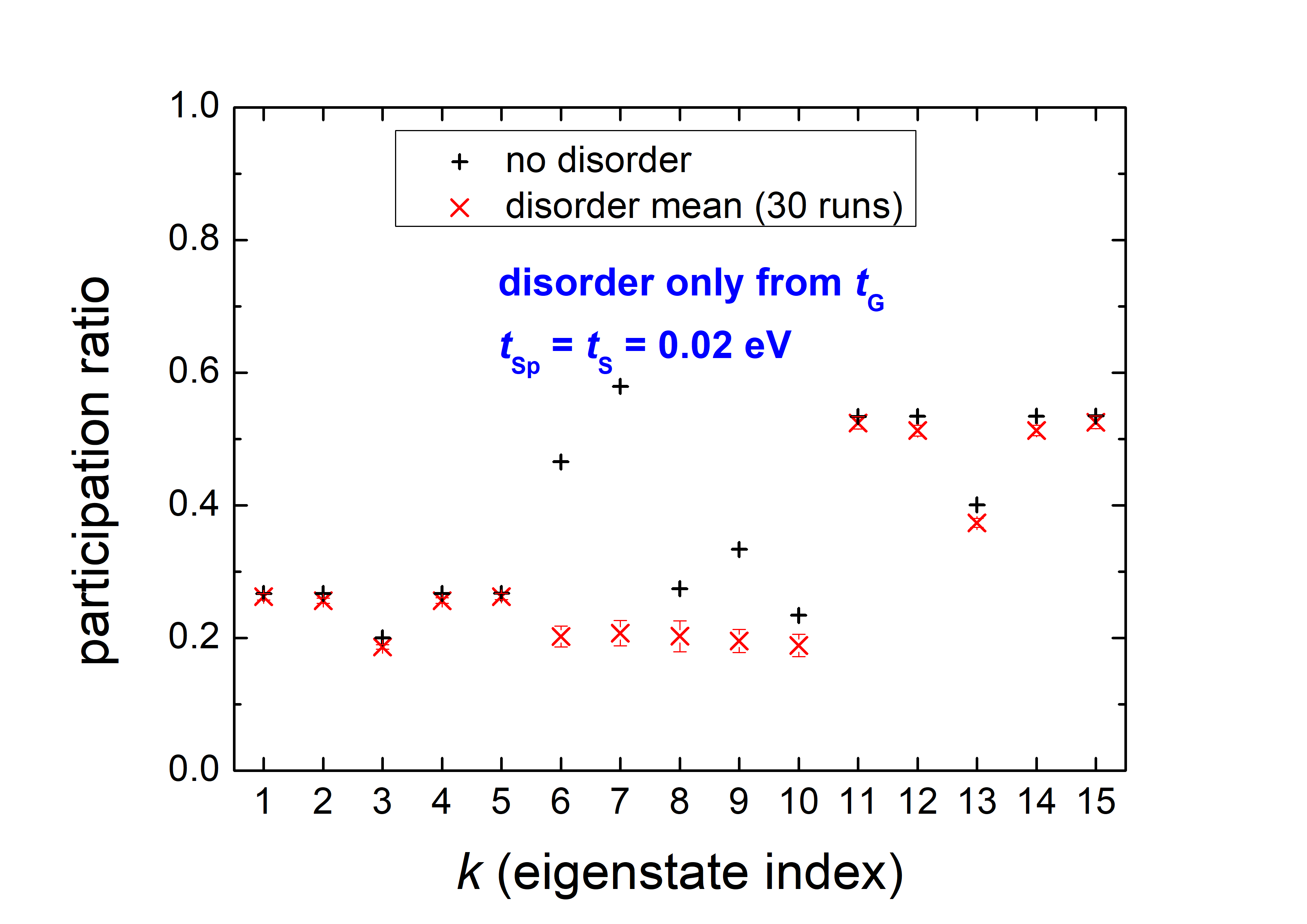} 
\includegraphics[width=0.48\textwidth]{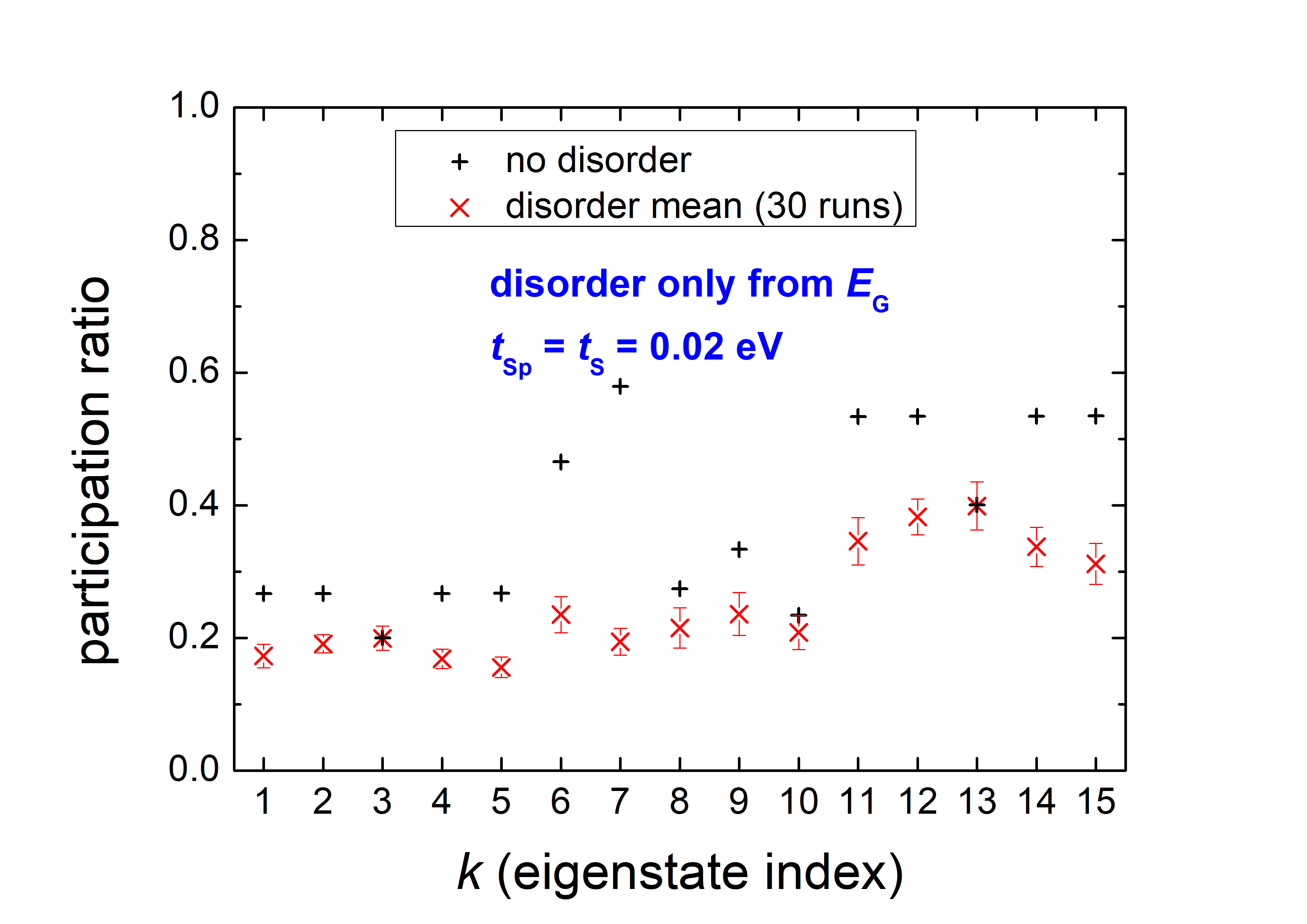}  
\caption{TB Fishbone Wire Model. G$_5$ sequence. $t_\mathrm{S} = t_\mathrm{S'} = 0.02$ eV. Effect of disorder, emanating from various separate sources, on Participation Ratio. 1st panel: disorder only from $t_\mathrm{S}$, 2nd panel: disorder only from $E_\mathrm{S}$, 3rd panel: disorder only from $t_\mathrm{G}$, 4th panel: disorder only from $E_\mathrm{G}$.
Error bars are also shown.}
\label{fig:PR+disorder-SeparateSources-tStS}
\end{figure}

\begin{figure}
\centering
\includegraphics[width=0.48\textwidth]{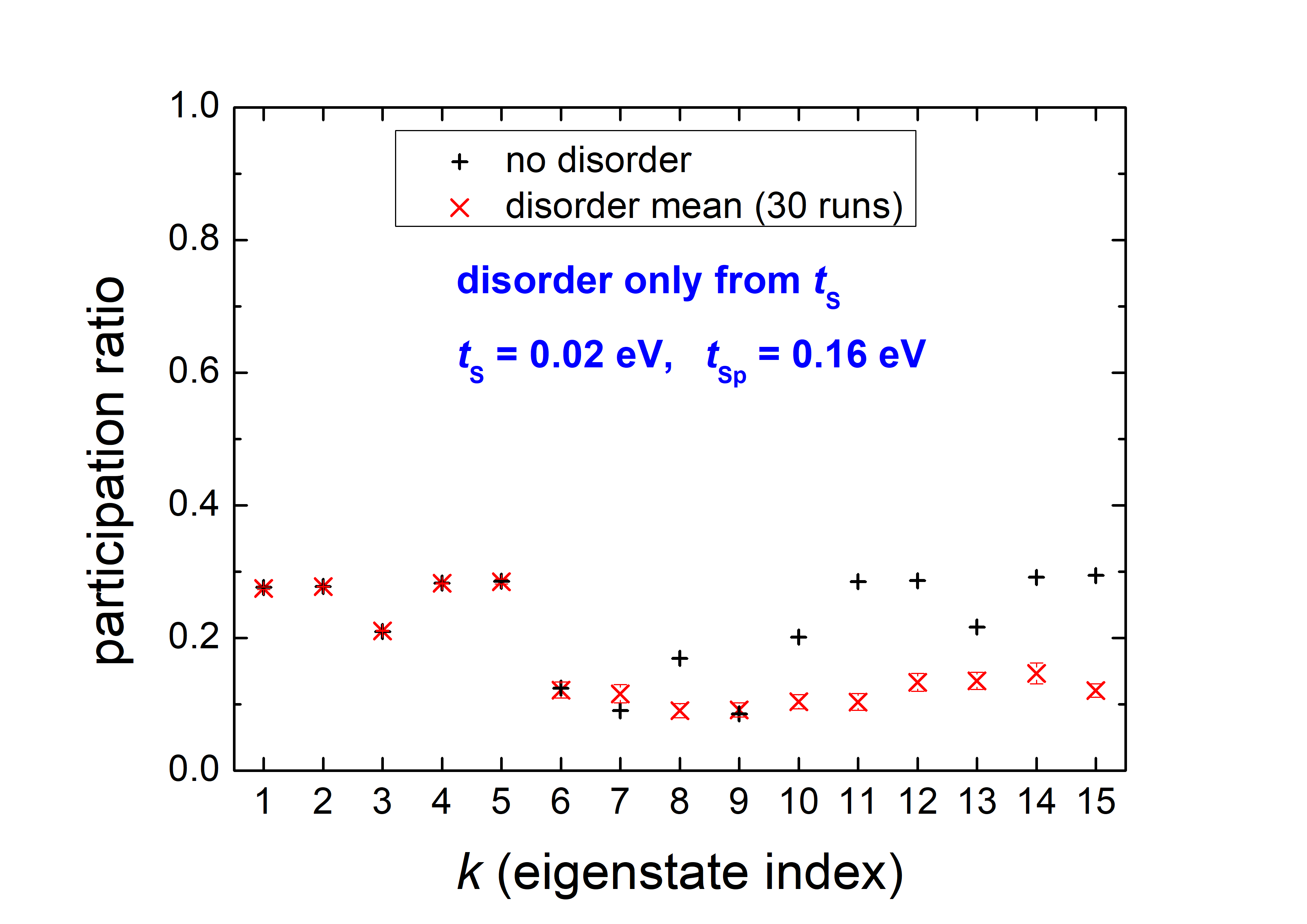} 
\includegraphics[width=0.48\textwidth]{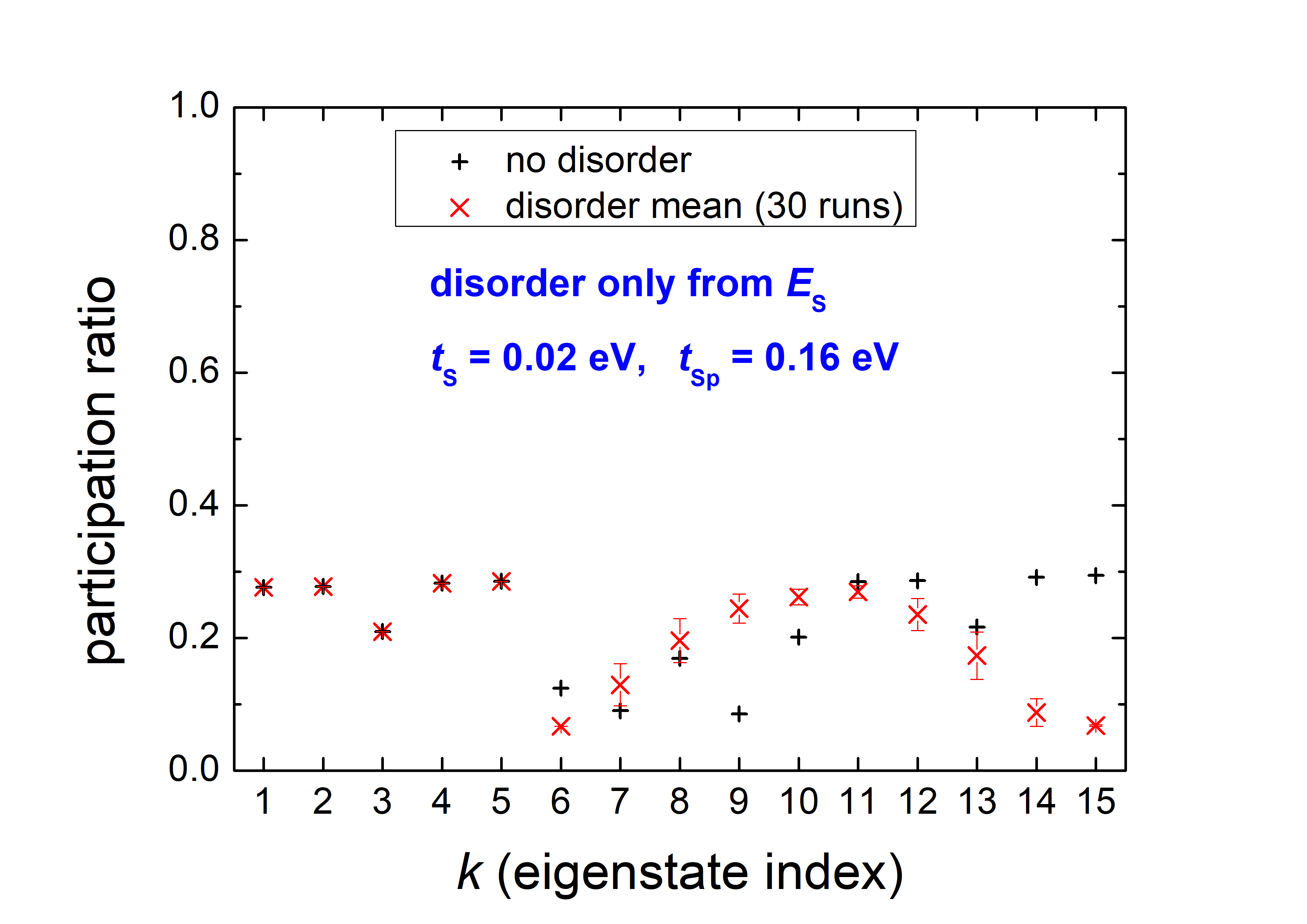} 
\includegraphics[width=0.48\textwidth]{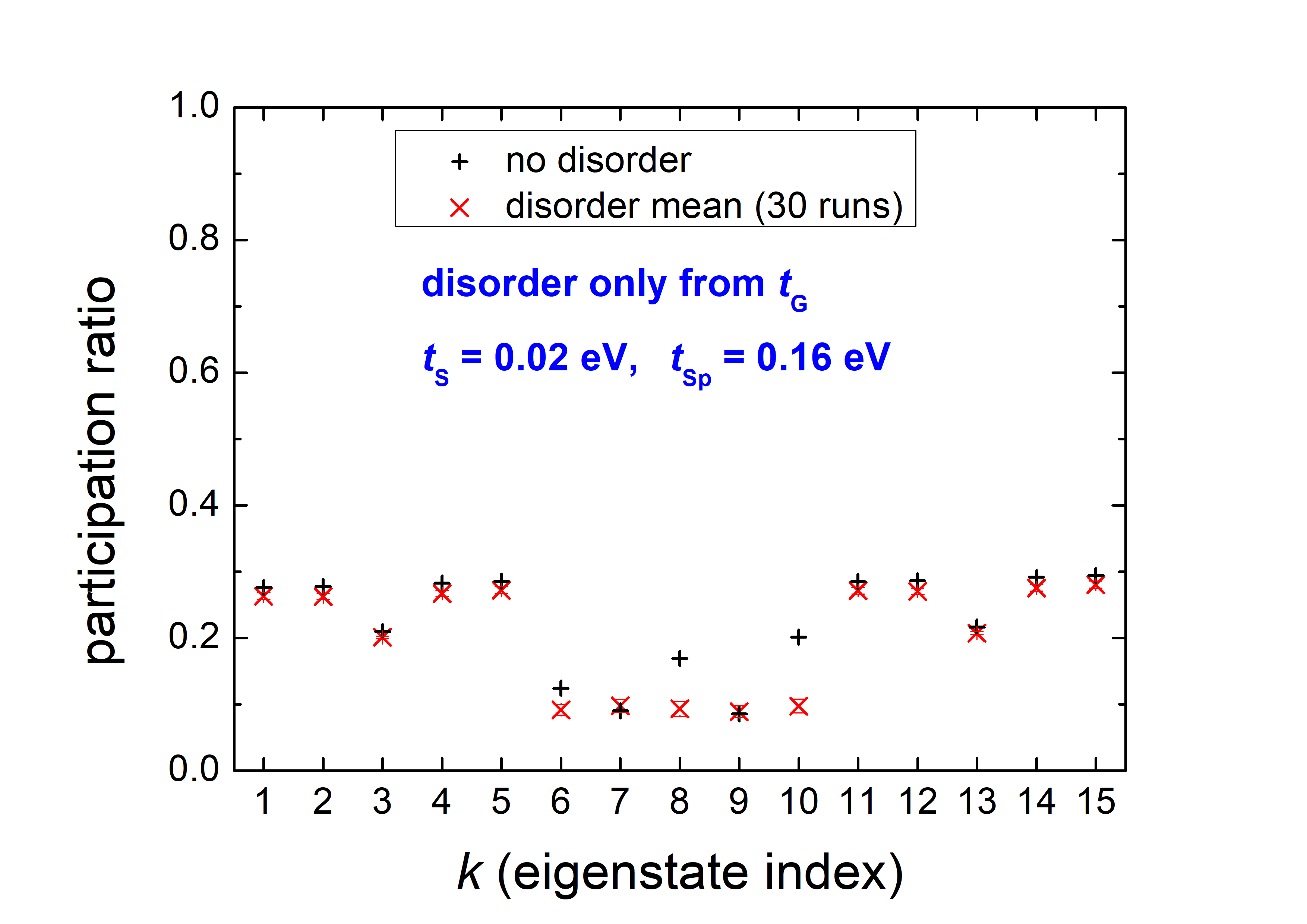} 
\includegraphics[width=0.48\textwidth]{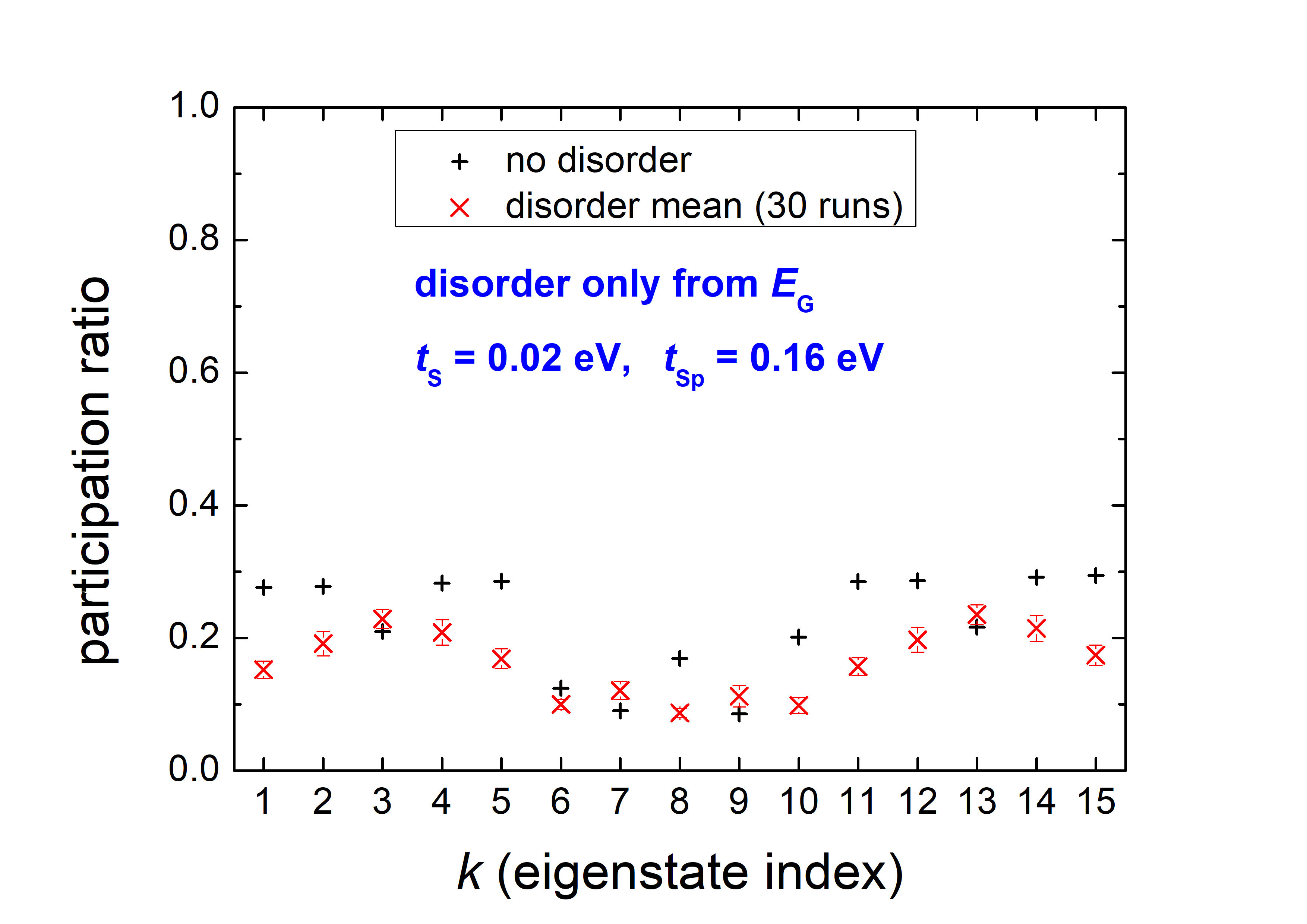} 
\caption{TB Fishbone Wire Model. G$_5$ sequence. 
$t_{\mathrm{S}} = 0.02$ eV $<$ $t_{\mathrm{S'}} = 0.16$ eV.	
Effect of disorder, emanating from various separate sources, on Participation Ratio. 1st panel: disorder only from $t_\mathrm{S}$, 2nd panel: disorder only from $E_\mathrm{S}$, 3rd panel: disorder only from $t_\mathrm{G}$, 4th panel: disorder only from $E_\mathrm{G}$. When we refer to disorder on $t_{\mathrm{S}}$, the random number generator is used in all $tS$ and $tSp$. Error bars are also shown.}
\label{fig:PR+disorder-SeparateSources-tStSp}
\end{figure}
	

\clearpage

\bibliographystyle{unsrtnat}
\bibliography{references}

\end{document}